\documentclass[a4paper,fleqn,usenatbib]{mnras}

\usepackage{newtxtext,newtxmath}

\usepackage[T1]{fontenc}
\usepackage{ae,aecompl}

\usepackage{graphicx}	
\usepackage{amsmath}	
\usepackage{amssymb}	

\title[Photospheric survey of three white dwarfs]{A far-UV survey of 
three hot, metal-polluted white dwarf stars: WD0455-282, WD0621-376, and
WD2211-495}

\author[Preval et al.]{Simon P. Preval,$^{1}$\thanks{E-mail: spp11@leicester.ac.uk}
Martin A. Barstow,$^{1}$
Matthew Bainbridge,$^{1}$
Nicole Reindl,$^{1}$
\newauthor
Thomas Ayres,$^{2}$
Jay B. Holberg,$^{3}$
John D. Barrow,$^{4}$
Chung-Chi Lee$^{4}$
John K. Webb,$^{5}$
\newauthor
and Jiting Hu,$^{5}$
\\
$^{1}$University of Leicester, Physics Building, University Road, 
Leicester, LE1 7RH, United Kingdom\\
$^{2}$Center for Astrophysics and Space Astronomy, University of 
Colorado, Boulder, CO 80309-0389, USA\\
$^{3}$Lunar and Planetary Laboratory, Sonett Space Sciences Building, 
University of Arizona, Tucson, AZ 85721, USA\\
$^{4}$DAMTP, Centre for Mathematical Sciences, University of Cambridge, 
Wilberforce Road, Cambridge CB3 0WA, UK\\
$^{5}$School of Physics, University of New South Wales, Sydney, 
NSW 2052, Australia
}

\date{Accepted Soon. Received YYY; in original form ZZZ}

\pubyear{2019}

\begin{document}
\label{firstpage}
\pagerange{\pageref{firstpage}--\pageref{lastpage}}
\maketitle

\begin{abstract}
Using newly obtained high-resolution data ($R\sim{1\times{10}^{5}}$) 
from the \textit{Hubble Space Telescope}, and archival UV data from 
the \textit{Far Ultraviolet Spectroscopic Explorer} we have conducted 
a detailed UV survey of the three hot, metal-polluted white dwarfs 
WD0455-282, WD0621-376, and WD2211-495. Using bespoke model atmospheres 
we measured $T_{\mathrm{eff}}$, log $g$, and photospheric abundances for 
these stars. In conjunction with data from Gaia we measured masses, 
radii, and gravitational redshift velocities for our sample of objects. 
We compared the measured photospheric abundances with those predicted by
radiative levitation theory, and found that the observed Si abundances 
in all three white dwarfs, and the observed Fe abundances in WD0621-376 
and WD2211-495, were larger than those predicted by an order of 
magnitude. These findings imply not only an external origin 
for the metals, but also ongoing accretion, as the metals not supported 
by radiative levitation would sink on extremely short timescales.
We measured the radial velocities of several absorption features along 
the line of sight to the three objects in our sample, allowing us to 
determine the velocities of the photospheric and interstellar components 
along the line of sight for each star. Interestingly, we made detections 
of circumstellar absorption along the line of sight to WD0455-282 with 
three velocity components. To our knowledge, this is the first such 
detection of multi-component circumstellar absorption along the line of 
sight to a white dwarf.
\end{abstract}




\section{Introduction}
White dwarf stars are ubiquitous throughout the galaxy, and mark the 
final stage of evolution for main sequence stars with masses 
$<8-10M_\odot$ \citep{iben1997a}. As they no longer produce heat through nuclear fusion, 
white dwarfs spend the rest of their lives cooling to background 
temperature via the emission of radiation. Through the use of calculated 
evolution tables, the temperature of a white dwarf can be effectively 
used as a means to measure how long the star has been cooling. 
Approximately 30-50\% of white dwarf stars are thought to have 
atmospheres polluted with metals \citep{barstow2014a,koester2014a} (B14 
and K14 hereafter). The mechanisms by which this pollution occurs is 
relatively well understood in the case of cool white dwarfs. The 
diffusion timescales for metals in these cases can be on the order of days
for hydrogen-rich (DA) white dwarfs with effective temperatures 
($T_{\mathrm{eff}}$) ranging between 12,000$<T_{\mathrm{eff}}<$20,000K 
\citep{paquette1986a}. Therefore, for metals to be observed in the 
atmosphere, the reservoir of material has to be continually replenished 
through accretion. However, in the case of hot white dwarfs 
($T_{\mathrm{eff}}>$20,000K) the physics becomes more complex. In this 
case, the radiative forces in the atmosphere are efficient enough to 
overcome the gravitational stratification, resulting in metals such as
C, N, O, and heavier being present in the photosphere through a process 
known as radiative levitation. In addition, as with the cool white 
dwarfs, accretion can also occur. Therefore, the observed photospheric 
abundance patterns in hot white dwarf stars will be a delicate balance 
between the flow of material sinking under gravity, the resistance to 
this flow from radiative levitation, and the introduction 
of additional material from accretion. Attempts have been made by 
\cite{chayer1994a,chayer1995a,chayer1995b} (Ch94/95 hereafter) to use 
radiative levitation theory to predict the observed abundance patterns 
in white dwarf stars, however, it has often been the case that these 
predictions bear little resemblance to what is observed. This is a 
consequence of the assumptions made. In particular, Ch94/95 assumed that 
the reservoir of material available to be levitated was effectively 
infinite.

The evidence for accretion into the atmospheres of white dwarf stars is 
compelling. A study by K14 analysed the atmospheres of 85 white dwarf 
stars with $17,000<T_{\mathrm{eff}}<27,000$K. From this sample, 48 out 
of the 85 (56\%) white dwarfs analysed were found to have signatures of 
metals in their atmospheres. Of these 48 metal-polluted white dwarfs, 
the photospheric compositions of 25 of these stars could be explained by 
radiative levitation alone. However, for the other 23 stars, K14 
concluded that active accretion was taking place to account for the 
observed abundance patterns. B14 conducted a survey of 89 DA white 
dwarfs observed by the Far Ultraviolet Spectroscopic Explorer (FUSE) 
with $16,000<T_{\mathrm{eff}}<77,000$K. In this sample, 33 out of 89 
white dwarfs were found to have atmospheres polluted with metals. In an 
attempt to understand the nature of this metal pollution, the authors
compared the measured photospheric abundances to those predicted by 
Ch94/95. Out of the 33 white dwarfs considered, 12 were 
found to have enhanced Si abundances relative to the predicted radiative 
levitation abundances, implying an external origin. Further 
investigating this, B14 looked at the ratios of abundances for each 
star, such as Si/C, Si/P, and Si/S. The authors then compared these 
ratios to those seen in bulk-Earth material, in CI chondrites, and in 
the sun. It was shown that the white dwarf Si/C ratios were similar to 
those observed in CI chondrite material, further implying that metal 
pollution in hot white dwarf stars originates from rocky bodies. 

In the case of cooler white dwarfs ($T_{\mathrm{eff}}<20,000$K) where
radiative levitation is inefficient, comparing photospheric abundance 
ratios such as C/O to those observed in other types of media allows the 
origin of the material to be inferred. \cite{wilson2016a} measured the 
photospheric compositions of 16 white dwarf stars with 
$17,000<T_{\mathrm{eff}}<23,000$K and calculated the C/O ratio for each 
object. The authors found that four of these objects had C/O ratios 
coincident with that observed in bulk-Earth material,
and five of these objects had C/O ratios coincident with that observed
in chondrite material. Other authors have also used this technique to 
study the compositions of accreted material such as 
\cite{kawka2016a,zuckerman2011a,xu2014a}. As well as photospheric 
abundance measurements for cool white dwarfs, observational data has 
revealed the presence of dusty debris disks orbiting white dwarf stars. 
An excellent example of this is the metal polluted white dwarf 
SDSS 1228+1040. Using time-resolved spectroscopic data from the William 
Herschel Telescope \cite{gansicke2006a} detected emission features from 
the Ca {\sc ii} triplet with laboratory wavelengths 8498, 8542, and 
8662\AA. The authors also found this emission did not vary over 
timescales of 20 minutes to one day. The authors concluded that this 
emission arose from a gaseous circumstellar disk around the white dwarf. 
More exotic cases of debris disks have been observed, such as that of 
WD1145+017. Using high cadence photometry, \cite{vanderburg2015a} made 
the surprising discovery that not only did WD1145+017 have a debris 
disk, this disk was found to be composed of several disintegrating 
objects. As this debris transits the white dwarf, it causes the star's 
brightness to dim by up to 40\%. Further studies such as those by 
\cite{rappaport2016a,gansicke2016a,xu2016a,croll2017a} have focused on 
characterising the type and compositions of the objects orbiting 
WD1145+017. Most recently, new observations by \cite{manser2019a} of
SDSS 1228+1040 have revealed the presence of an Fe-core planetesimal 
orbiting the white dwarf. 

The extensive literature concerning the accretion of metals onto cool
white dwarf stars is compelling. However, the question of the origin of
metal pollution in hot white dwarfs is still an open question. In this 
paper we present our photospheric and interstellar survey of three hot, 
DA white dwarfs stars, namely WD0455-282, WD0621-376, and WD2211-495. 
First, we describe new Hubble Space Telescope observations obtained for 
the three white dwarfs, as well as archival 
\emph{Far Ultraviolet Spectroscopic Explorer} (FUSE) data. We then 
describe the model atmosphere calculations performed, and how these were 
used to measure the effective temperature, surface gravity, and 
photospheric compositions of the stars. We then conduct a line survey of 
the three objects, and identify the velocity populations present along 
the line of sight to the white dwarfs. We then discuss the results 
obtained, and consider the photospheric compositions of the white dwarfs 
in the context of radiative levitation. We also discuss the results of 
the interstellar survey. Lastly, we make concluding remarks, and detail
the future work planned.

\section{Observations}\label{obsec}
\begin{figure*}
\begin{centering}
\includegraphics[width=180mm]{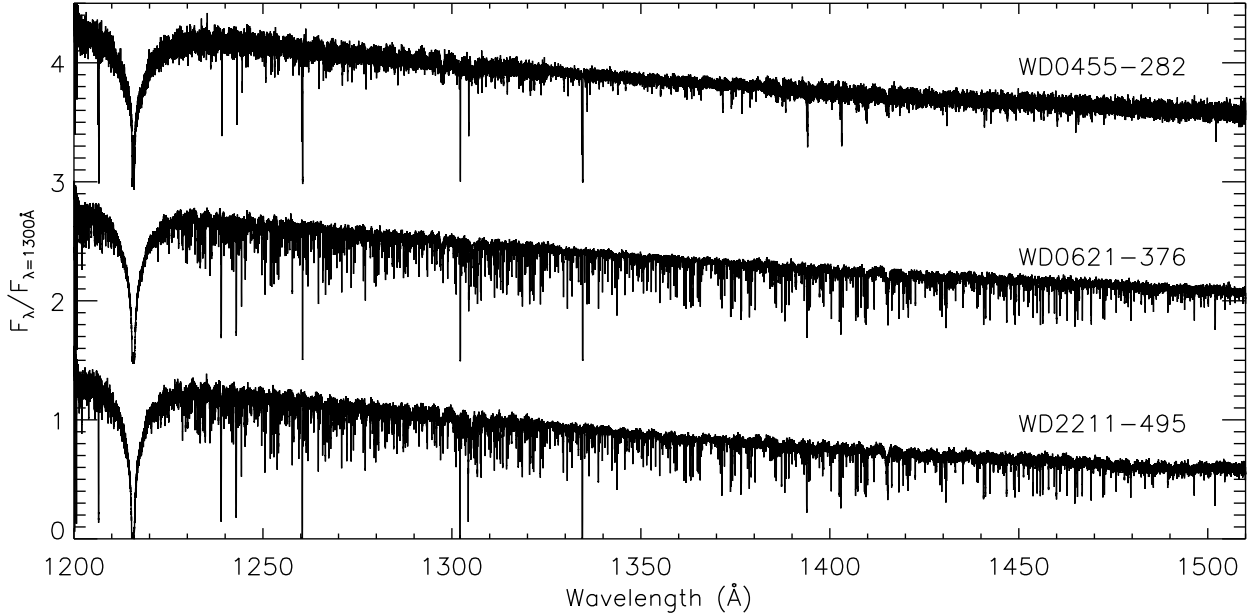}
\caption{STIS spectra obtained for WD0455-282, WD0621-376, and 
WD2211-495. The spectra have been normalised by the flux at 1300\AA, and 
offset from each other for clarity.}
\label{fig:stisspec}
\end{centering}
\end{figure*}

WD0455-282, WD0621-376, and WD2211-495 were observed using the Space 
Telescope Imaging Spectrometer (STIS) aboard the 
\emph{Hubble Space Telescope} as part of program GO-14791. Each star was 
observed using the E140H FUV/MAMA echelle grating (nominal resolution 
$R= {\lambda}/{\Delta\lambda}\sim 1{\times}10^{5}$, 
\citealt{hernandez2012a}), with a 0.2x0.2 aperture. The aperture size 
was chosen so as to maximise the flux throughput. Further, special, 
deeper than normal, exposures of the wavelength 
calibration lamp were taken to aid in the assignment of precise 
wavelength scales. The STIS echellegram reduction and co-addition 
procedures we adopted have been fully described in a 
previous paper \citep{hu2019a}. In brief, the raw spectral datasets 
were processed through a version of the STIS pipeline with a number of 
specially updated reference files, including new dispersion solutions 
for the specific grating settings used. The individual echelle orders of 
the processed echellegrams were then concatenated, 
imposing an active blaze correction to ensure optimum flux continuity 
across the overlapping zones of the adjacent orders. If multiple 
sub-exposures were taken in a given grating setting, these were co-added 
after aligning relative to the leading spectrum by cross-correlation.  
Finally, the overlapping regions between the two separate grating 
settings (1307\AA\, and 1416\AA) were adjusted to match the fluxes and 
again aligned in wavelength by cross-correlation. The alignment shifts 
typically were small, often only 0.1 km s$^{-1}$ in magnitude. The 
final wavelength coverage for the spliced spectra was approximately 
1200$-$1520\AA. The average signal-to-noise ratio per 
resolution element ($\sim$3 km s$^{-1}$) in the co-added spectra ranged 
from about 35 (WD0455-282) to 60--70 (WD2211-495 and WD0621-376, 
respectively). We summarise the STIS observations for each star in 
Table \ref{table:stisobs}, and plot the STIS spectra of each object in 
Figure \ref{fig:stisspec}. 

\begin{table*}
\centering
\caption{List of spectral datasets obtained from the HST for each object 
using STIS with the E140H grating. $\lambda_{c}$ is the central 
wavelength of the exposure. The start time is given in the Universal 
Time Zone (UT).}
\begin{tabular}[H]{@{}lllcc}
\hline
Object & Observation ID & Start Time/Date & $\lambda_{c}$ (\AA) & Exposure time (s) \\
\hline
WD0455-282 & OD7QD0IUQ & 05:43:21 15/08/2017 & 1307 & 2487.0 \\
           & OD7QD0JDQ & 07:15:40 15/08/2017 & 1307 & 2964.0 \\
           & OD7QD0JHQ & 08:51:06 15/08/2017 & 1307 & 2964.0 \\
           & OD7QD1010 & 10:29:50 15/08/2017 & 1416 & 2487.0 \\
           & OD7QD1020 & 12:02:00 15/08/2017 & 1416 & 2964.0 \\
           & OD7QD1030 & 13:37:26 15/08/2017 & 1416 & 2964.0 \\
WD0621-376 & OD7QC0H5Q & 02:40:08 10/10/2017 & 1307 & 2525.0 \\
           & OD7QC0HRQ & 04:04:42 10/10/2017 & 1307 & 3002.0 \\
           & OD7QC0030 & 05:42:08 10/10/2017 & 1416 & 2881.0 \\
           & OD7QC0030 & 07:15:33 10/10/2017 & 1416 & 3002.0 \\
WD2211-495 & OD7QA0MRQ & 05:49:44 15/10/2017 & 1307 & 2610.0 \\
           & OD7QA0020 & 07:19:07 15/10/2017 & 1416 & 2966.0 \\
\hline
\end{tabular}
\label{table:stisobs}
\end{table*}

\begin{figure*}
\begin{centering}
\includegraphics[width=180mm]{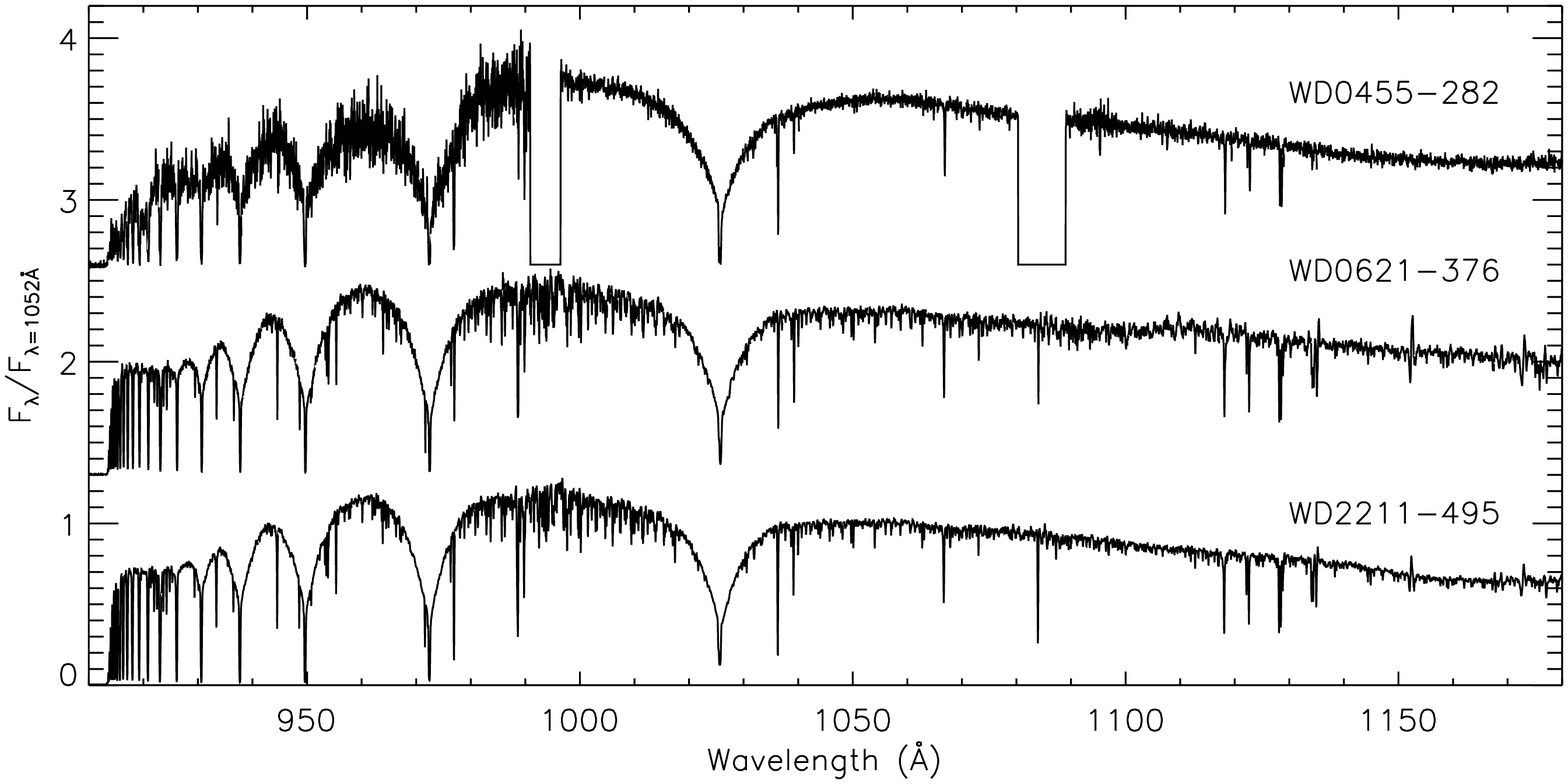}
\caption{FUSE spectra obtained for WD0455-282, WD0621-376, and 
WD2211-495. The spectra have been normalised by the flux at 1052\AA, and 
offset from each other for clarity.}
\label{fig:fusespec}
\end{centering}
\end{figure*}

To complement our STIS observations, and provide a means with which to 
measure the effective temperature ($T_{\mathrm{eff}}$) and surface 
gravity (log $g$) for the three objects, we constructed high 
signal-to-noise spectra using archival observations from the 
\emph{Far Ultraviolet Spectroscopic Explorer} (FUSE). We summarise the 
FUSE exposures used in this work in Table \ref{table:fuseobs}, and plot 
the FUSE spectra for each object in Figure \ref{fig:fusespec}. Unlike 
STIS, which uses echelle-blaze optics, FUSE used four individual mirrors 
to reflect light onto two detectors, resulting in eight individual 
spectra (see \citealt{moos2000a}). Collectively, these exposures spanned 
905-1184\AA, and encompassed the H-Lyman series from Ly-$\beta$, down to 
the lyman limit ($\sim{911}$\AA). There were three apertures available 
to use. These were the LWRS (low resolution, 30'x30'), MDRS (medium 
resolution, 4'x20'), and HIRS (high-resolution, 1.25'x20') apertures 
with resolutions ranging from 15,000 to 21,000. After FUSE launched, it 
became evident that there were distortions present in the optical 
design, caused by heating and cooling of the individual components. This 
resulted in the eight spectra having different wavelength shifts. In 
addition, these distortions meant that the object being observed often 
drifted out of the field of view when the MDRS and HIRS apertures were 
used. This resulted in the flux of the observed spectra being attenuated 
by arbitrary amounts. Therefore, when constructing the spectra for this 
work, we only included observations where the LWRS aperture was used 
where possible. In the case of WD0455-282 where MDRS observations were 
used, we omitted exposures where the flux was attenuated, or where there 
was no flux at all.

We note that, in addition to the FUSE and HST datasets, there also 
exists UV spectroscopy from the International Ultraviolet Explorer 
(IUE), with a resolution $\sim{1\times{10}^4}$. IUE data for the three
objects considered in this paper were analysed by \cite{holberg1998a},
who catalogued the various absorption features observed in their 
respective spectra. We opt not to use these IUE data in favour of the 
higher resolution FUSE and STIS data described hitherto.

\begin{table*}
\centering
\caption{List of spectral datasets obtained from FUSE for each
object. The start time is given in theUniversal Time Zone (UT).}
\begin{tabular}[H]{@{}lccccc}
\hline
Object & Observation ID & Exposures & Start Time/Date & Aperture & Exposure time (s) \\
\hline
WD0455-282 & P10411010 & 15 & 02:14:28 03/02/2000 & MDRS & 19667.0 \\
           & P10411020 &  8 & 05:07:49 04/02/2000 & MDRS & 10122.0 \\
           & P10411020 & 13 & 09:36:51 07/02/2000 & MDRS & 17675.0 \\
WD0621-376 & P10415010 & 19 & 05:28:58 06/12/2000 & LWRS &  8371.0 \\
WD2211-495 & M10303020 &  5 & 17:18:58 21/10/1999 & LWRS &  2720.0 \\
           & M10303030 &  9 & 19:43:40 25/10/1999 & LWRS &  4160.0 \\ 
           & M10303040 &  7 & 21:45:44 01/11/1999 & LWRS &  2976.0 \\ 
           & M10303050 & 11 & 06:55:16 03/06/2000 & LWRS &  5157.0 \\
           & M10303060 &  8 & 05:15:34 29/06/2000 & LWRS &  4193.0 \\
           & M10303070 & 12 & 20:43:02 17/08/2000 & LWRS &  5260.0 \\
           & M10303080 &  9 & 06:09:32 24/10/2000 & LWRS &  3954.0 \\
           & M10303090 & 11 & 09:29:25 24/10/2000 & LWRS &  5025.0 \\
           & M10303100 &  9 & 12:49:19 24/10/2000 & LWRS &  5795.0 \\
           & M10303110 & 13 & 05:28:46 25/10/2000 & LWRS &  5877.0 \\
           & M10303120 & 11 & 10:28:36 25/10/2000 & LWRS &  5430.0 \\
           & M10303130 &  8 & 04:07:07 01/08/2002 & LWRS &  3796.0 \\
           & M10303140 &  7 & 14:23:40 10/06/2002 & LWRS &  3774.0 \\
           & M10303150 & 10 & 17:20:00 25/09/2002 & LWRS &  5460.0 \\
           & M10303160 & 13 & 23:28:41 24/05/2002 & LWRS &  6929.0 \\
           & M10303180 & 12 & 04:14:39 17/09/2003 & LWRS &  5875.0 \\
           & M10315010 &  7 & 18:26:46 07/05/2004 & LWRS &  3699.0 \\
           & M10315040 &  7 & 17:27:33 06/07/2004 & LWRS &  3889.0 \\
\hline
\end{tabular}
\label{table:fuseobs}
\end{table*}

\section{Stellar atmosphere models}
All model atmospheres described in this paper were calculated using the 
non-LTE model atmosphere code {\sc tlusty}, version 201 
\citep{hubeny1988a,hubeny1995a}. Synthetic spectra were then generated 
from these model atmospheres using {\sc synspec} \cite{hubeny2011a}. 
These spectra spanned 900-1600\AA, and were convolved with an 
instrumental Gaussian profile whose FWHM (full-width, half maximum) was
dependent upon the observational data being considered. For FUSE data, 
the FWHM was set to 0.0641\AA, and for the STIS data the FWHM was 
calculated as $\lambda/114,000$ as given in the STIS instrument manual. 
The H-Lyman absorption profiles were calculated using the Stark 
Broadening tables of \cite{tremblay2009a}. These tables are an updated 
version of the tables presented in \cite{lemke1997a}, including 
non-ideal effects such as those resulting from perturbations to the 
absorber due to protons and electrons.

Given the high temperature of the white dwarf stars considered, we have 
created new model ions for C, N, O, Si, P, and S to account for the 
presence of higher ionization states. For the model atmospheres 
calculated, we include model ions for C {\sc i-v}, N {\sc i-vi}, 
O {\sc i-vii}, Si {\sc i-viii}, P {\sc i-viii}, S {\sc i-viii}, 
Fe {\sc iii-vii}, and Ni {\sc iii-vii}. In addition, we also include 
single-level ions for C {\sc vi}, N {\sc vii}, O {\sc viii}, 
Si {\sc ix}, P {\sc ix}, S {\sc ix}, Fe {\sc viii}, and Ni {\sc viii}.

Full line-blanketing effects for $>9,000,000$ transitions of Fe and Ni 
are accounted for via the opacity sampling method. For Fe, the atomic 
data including energy levels and transitions originates from 
\cite{kurucz1992a}, and the photoionization cross sections (PICS) from 
the Opacity Project \citep{seaton1994a}. For Ni, the atomic data also 
originates from \cite{kurucz1992a}, and the PICS for Ni {\sc vii} were 
calculated using an hydrogenic approximation. For Ni {\sc iv-vi} we use 
the PICs calculated by \cite{preval2017a}.

\section{Methodology}
\subsection{General photospheric parameter measurement}\label{abnsec}
It is currently untenable to calculate model atmosphere grids that can 
account for variations in $T_{\mathrm{eff}}$, log $g$, and all of the 
individual abundances simultaneously. To do so would require the 
calculation of thousands, if not millions, of stellar atmospheric models
that incorporate the various possible permutations of these variables. 
Therefore, for each star, we adopted an iterative procedure to determine 
the best fitting values of these parameters. First, we calculated a 
starting NLTE model atmosphere with {\sc tlusty} using values of 
$T_{\mathrm{eff}}$, log $g$, and metal abundances from 
\cite{barstow2003b} and B14. Then, for each metal abundance we 
wanted to measure, we used {\sc synspec} to calculate a grid of model 
atmospheres with different abundances by stepping away from the starting 
model in LTE. We then measured the abundances using these grids. Once we 
measured the abundances, we used {\sc tlusty} to calculate an NLTE model 
atmosphere grid with these abundances, varying in $T_{\mathrm{eff}}$ and 
log $g$. We then measured $T_{\mathrm{eff}}$ and log $g$ using the 
H-Lyman lines. After measuring $T_{\mathrm{eff}}$ and log $g$, we 
calculated an NLTE model atmosphere using the measured 
$T_{\mathrm{eff}}$, log $g$, and metal abundances. We then repeated the 
aforementioned steps until the measured $T_{\mathrm{eff}}$, log $g$, and 
metal abundances were converged. Our convergence criterion was defined 
such that the difference between the parameters measured in the previous 
and current iterations were less than their measurement uncertainties. 
In total, we performed seven iterations for WD0455-282, and six 
iterations for WD0621-376 and WD2211-495. Below, we describe how the 
metal abundances, $T_{\mathrm{eff}}$, and log $g$ were measured in more 
detail.

\subsection{$T_{\mathrm{eff}}$, log $g$, and abundance measurements}
In Table \ref{table:trans} we list the transitions used to measure the 
abundances of C, N, O, Si, P, S, Fe, and Ni. Fe and Ni are difficult 
metals for which to measure an abundance. This is because the atomic 
data can be quite uncertain with respect to their centroid wavelengths, 
broadening parameters, and oscillator strengths. In an attempt to 
mitigate this, we used 8--10 Fe {\sc v} and Ni {\sc v} transitions, and 
simultaneously fit all of these lines to find the best 
abundance value. We note that for our abundance determinations we use
only one ionization state per ion. Generally, wherever possible, we have
tried to use lines arising from transitions between excited states. This
is because resonant transitions have been known to be accompanied by 
high ionization state ISM lines (also termed ``circumstellar'', see 
\citealt{bannister2003a,dickinson2012b}) which can potentially 
compromise the abundance measurement. We discuss the lines used, and the 
reason why they were used here. In the case of C, only the C {\sc iii} 
transitions were visible in the spectra considered. For N,
the N {\sc iv} multiplet is present at 912-915\AA, however, these lines
are in a region of spectrum where the signal-to-noise is sub optimal. In
addition, the multiplet is contaminated by interstellar H. Therefore, we
opt to use only the N {\sc v} doublet. For O there are three O {\sc iv}
lines and one O {\sc v} line ($\sim{1371}$\AA) that can be used. In this
case, we opted to use the O {\sc iv} lines simply because there were 
more lines to use. For Si the Si {\sc iii} ionization state has a low
population relative to Si {\sc iv} due to the temperatures of the 
objects considered ($>60,000$K). For P, the P {\sc v} ionization state
is the most populated relative to P {\sc iv}, again due to the 
temperatures of the objects considered. For S, the S {\sc vi} resonant
transitions are detected at 933\AA\, and 945\AA. However, these lines 
are in close proximity to Ly-$\delta$ and Ly-$\epsilon$, meaning they
are sensitive to the wings of these H lines. In addition, abundance 
measurements made using these S {\sc vi} lines have been shown to be in 
great disagreement with measurements made using S {\sc iv} and S{\sc v}
(see, for example, \citealt{preval2013a}). Lastly, for Fe and Ni, the 
strongest lines come from Fe {\sc v} and Ni {\sc v}. Only a few 
Fe/Ni {\sc vi} lines are visible in the spectrum that could potentially 
be used. Therefore, we opted to use the Fe/Ni {\sc v} lines due to the 
number of these lines available.

The STIS spectra used in this work have $>$50,000 data points. Reading 
in, and fitting all of these data points is very computationally 
demanding. Therefore, we opted to fit individual transitions by 
extracting a section of spectrum encompassing the feature we want to 
measure. The wavelength regions extracted for each transition were the 
same for all three stars, and are listed in Table \ref{table:trans}. To 
measure the individual metal abundances, we used the X-Ray spectral 
analysis package {\sc xspec} \citep{arnaud1996a}, which uses a 
$\chi^{2}$ minimisation procedure to interpolate a model grid to the 
observed data, and determine the best fitting values for these 
parameters with the reduced $\chi^{2}$ value ($\chi^{2}_{\mathrm{red}}$)
being as close to unity as possible. To ensure we obtained the best 
fitting abundance values, we calculated $\chi^{2}_{\mathrm{red}}$ over 
all abundance values in the model grid. 

\begin{table}
\centering
\caption{List of transitions used to measure the photospheric abundances 
of the three white dwarfs considered. Transitions with 
wavelengths in the STIS range (1200-1520\AA) were used to measure the 
photospheric velocity.}
\begin{tabular}[H]{@{}lccc}
\hline
Ion & Wavelength (\AA) & log $gf$ & Extracted Region (\AA) \\
\hline
C {\sc iii} & 1174.933 & -0.468 & 1170--1180 \\
C {\sc iii} & 1175.263 & -0.565 & 1170--1180 \\
C {\sc iii} & 1175.590 & -0.690 & 1170--1180 \\
C {\sc iii} & 1175.711 &  0.009 & 1170--1180 \\
C {\sc iii} & 1175.987 & -0.565 & 1170--1180 \\
C {\sc iii} & 1176.370 & -0.468 & 1170--1180 \\
C {\sc iii} & 1247.383 & -0.314 & 1245--1255 \\
N   {\sc v} & 1238.821 & -0.505 & 1235--1245 \\
N   {\sc v} & 1242.804 & -0.807 & 1235--1245 \\
O  {\sc iv} & 1338.615 & -0.632 & 1335--1350 \\
O  {\sc iv} & 1342.990 & -1.333 & 1335--1350 \\
O  {\sc iv} & 1343.514 & -0.380 & 1335--1350 \\
Si {\sc iv} & 1122.485 &  0.220 & 1115--1130 \\
Si {\sc iv} & 1128.340 &  0.470 & 1115--1130 \\
Si {\sc iv} & 1393.755 &  0.030 & 1390--1405 \\
Si {\sc iv} & 1402.770 & -0.280 & 1390--1405 \\
P   {\sc v} & 1117.977 & -0.010 & 1115--1130 \\
P   {\sc v} & 1128.008 & -0.320 & 1115--1130 \\
S   {\sc v} & 1501.760 & -0.489 & 1495--1505 \\
Fe  {\sc v} & 1320.409 &  0.243 & 1315--1325 \\
Fe  {\sc v} & 1387.937 &  0.660 & 1385--1395 \\
Fe  {\sc v} & 1402.385 &  0.229 & 1390--1405 \\
Fe  {\sc v} & 1430.572 &  0.597 & 1425--1435 \\
Fe  {\sc v} & 1440.528 &  0.448 & 1440--1450 \\
Fe  {\sc v} & 1446.617 &  0.468 & 1440--1450 \\
Fe  {\sc v} & 1448.847 &  0.309 & 1440--1450 \\
Fe  {\sc v} & 1455.555 &  0.277 & 1450--1460 \\
Fe  {\sc v} & 1456.162 &  0.173 & 1450--1460 \\
Fe  {\sc v} & 1464.686 &  0.471 & 1460--1470 \\
Ni  {\sc v} & 1257.626 &  0.585 & 1255--1265 \\
Ni  {\sc v} & 1261.760 &  0.478 & 1255--1265 \\
Ni  {\sc v} & 1273.204 &  0.395 & 1265--1275 \\
Ni  {\sc v} & 1279.720 &  0.300 & 1275--1290 \\
Ni  {\sc v} & 1300.979 & -0.048 & 1295--1305 \\
Ni  {\sc v} & 1307.603 &  0.168 & 1305--1315 \\
Ni  {\sc v} & 1318.515 &  0.307 & 1315--1325 \\
Ni  {\sc v} & 1336.136 &  0.557 & 1335--1345 \\
Ni  {\sc v} & 1342.176 &  0.345 & 1335--1345 \\
\hline
\end{tabular}
\label{table:trans}
\end{table}

To measure the $T_{\mathrm{eff}}$ and log $g$ of each star, we first 
extracted wavelength regions from FUSE spectra encompassing four Lyman 
lines, namely Ly-$\beta$, Ly-$\gamma$, Ly-$\delta$, and Ly-$\epsilon$. 
Ly-$\alpha$ is present in the STIS data, however, this line cannot be 
used to measure $T_{\mathrm{eff}}$ and log $g$ due to contamination from 
interstellar H absorption. In the extracted spectra, we removed 
erroneous photospheric and interstellar absorption features. We then 
used XSPEC to interpolate our $T_{\mathrm{eff}}$/log $g$ model grids to 
the Lyman lines.

The statistical uncertainties on the abundances, $T_{\mathrm{eff}}$, and
log $g$ were determined by considering the change in $\chi^{2}$. For the 
abundances, the uncertainties were obtained by reading off the abundance 
values for which $\Delta\chi^{2}=1$ corresponding to a $1\sigma$ 
confidence interval for one variable parameter. It should 
be noted that the abundance uncertainties do not account for the 
uncertainty in the $T_{\mathrm{eff}}$ and log $g$ measurement. This is 
because in the iteration scheme, $T_{\mathrm{eff}}$ and log $g$ are held
fixed for abundance measurements. For $T_{\mathrm{eff}}$ and log $g$, 
the uncertainties were calculated by reading off the $T_{\mathrm{eff}}$ 
and log $g$ values for which $\Delta\chi^{2}=2.2957$ corresponding to a 
$1\sigma$ confidence interval for two variable parameters. The 
statistical uncertainties on quantities measured in this 
work are dependent upon the uncertainty on the observed flux. In the 
case of FUSE data, the uncertainty on the observed flux can be greatly 
underestimated due to calibration issues with the instruments \citep{moos2000a}. 
Therefore, to determine the uncertainties on quantities 
measured using FUSE data, a few, additional steps were required. First,
we fit the observed data by interpolating the model grids with 
{\sc xspec}. This results in a $\chi_{\mathrm{red}}^{2}$ value not equal 
to unity. Then, we expressed the flux uncertainties as a fractional 
percentage of the observed flux (e.g. 2.5\%). We then varied this 
fractional percentage so as to obtain a $\chi_{\mathrm{red}}^{2}$ of 
unity. We then determined the uncertainty on the relevant quantities by 
reading off the $T_{\mathrm{eff}}$/log $g$ values corresponding to 
$\Delta\chi^{2}=2.2957$, and the abundance values corresponding to 
$\Delta\chi^{2}=1.0$.

\subsection{Astrophysical \& astrometric quantities}
With the second data release from the astrometric observatory Gaia, the 
positions and parallaxes of hundreds of thousands of white dwarfs are 
now known to milli- and micro-arcsecond precision. Using parallax data 
information from Gaia, the observed STIS spectrum, and our best fitting 
synthetic spectrum, we can measure the radius $R_{*}$ of the 
white dwarf. The distance $D$ in parsecs to the white dwarf is 
calculated as
$\overline{\omega}$:
\begin{equation}
D=\frac{1000}{\overline{\omega}}.
\end{equation}
where $\overline{\omega}$ is the parallax in milliarcseconds. Then,
$R_{*}$ is calculated as
\begin{equation}\label{eqn:rad}
R_{*} = D \sqrt{\frac{F_{\lambda}}{4\pi{H}_{\lambda}}}
\end{equation}
where $F_{\lambda}$ is the observed flux, and $H_{\lambda}$ is the 
Eddington flux as obtained from the synthetic spectrum of the white 
dwarf in question. $F_{\lambda}$ and $H_{\lambda}$ are functions of 
wavelength. Therefore, we adopted a statistical approach to find the 
best value for $F_{\lambda}/H_{\lambda}$. Firstly, we extracted a region 
of spectrum from the synthetic and observed datasets where there were 
relatively few absorption features. We found that the region 
1480-1500\AA\, is suitable. We then calculated the mean value of 
$F_{\lambda}/H_{\lambda}$, weighted by their reciprocal squared errors. 
This value was then used to determine $R_{*}$. With $R_{*}$ and log $g$, 
we can then determine the mass of the white dwarf, $M_{WD}$, calculated 
as
\begin{equation}\label{eqn:mass}
M_{\mathrm{WD}}=\frac{gR_{*}^{2}}{G}
\end{equation}
where $G$ is the universal gravitational constant. Finally, with the 
mass and radius, we can then find the gravitational redshift 
$z_{\mathrm{gr}}$. In velocity units, the redshift, $V_{\mathrm{gr}}$, 
is calculated as
\begin{equation}
V_{\mathrm{gr}} = cz_{\mathrm{gr}} = \frac{GM_{*}}{cR_{*}}
\end{equation}
where $c$ is the speed of light. 

\subsection{Velocity measurements}
The exceptional calibration and resolution of STIS allows for the 
precise measurement of the photospheric (or radial) and interstellar 
velocities of the three white dwarfs. When measuring the abundances of 
the various metal absorption features, {\sc xspec} shifts the synthetic
spectrum in wavelength space to find the best fit to the data. To 
measure the photospheric velocity, we calculate the average velocity of
all the transitions in the STIS data listed in Table \ref{table:trans}.
The uncertainty in the velocity measurement is then determined by 
calculating the standard deviation of the velocities. 

Again, thanks to the high-resolution of the STIS spectra, it is possible 
to separate blended ISM features into their individual components. While 
a detailed study of these features quantifying column densities and 
temperatures is beyond the scope of this work, we do, however, measure 
the velocities of these features. To measure the velocities of the ISM 
lines we fit Gaussian profiles to the observed absorption features, 
parameterised by the wavelength centroid, the absorption depth, and the 
Doppler width. The exact parameterisation is given in the Appendix. The 
advantage to the profile used is that it can account for line 
saturation. The best fitting parameters were determined by using a 
$\chi^2$ minimization procedure as implemented in the {\sc idl} package 
{\sc mpfit} \citep{markwardt2009a}. The interstellar velocity was 
measured by calculating the average of the measured velocities of the 
lines listed in Table \ref{table:ismtrans}, weighted by their inverse 
square errors. The uncertainty on the weighted ISM component was 
determined by calculating the standard deviation of the measured 
velocities. When we found evidence of multiple component ISM absorption, 
we calculated the average velocities of lines with similar velocities. 
We note that there are ISM lines that exist in the FUSE spectra of our 
objects. However, we opted not to use these lines in our velocity 
measurements due to the lower-accuracy wavelength calibration in the 
FUSE spectra.

\begin{table}
\centering
\caption{List of transitions used to measure the ISM velocities along 
the line of sight to the white dwarfs.}
\begin{tabular}[H]{@{}lccc}
\hline
Ion & Wavelength (\AA) & log $gf$ \\
\hline
N    {\sc i} & 1200.22329 & -0.4589  \\
N    {\sc i} & 1200.70981 & -0.7625  \\
Si {\sc iii} & 1206.4995  &  0.2057  \\
Si  {\sc ii} & 1260.4221  &  0.3911  \\
O    {\sc i} & 1302.16848 & -0.6195  \\
Si  {\sc ii} & 1304.3702  & -0.7329  \\
C   {\sc ii} & 1334.5323  & -0.5912  \\
\hline
\end{tabular}
\label{table:ismtrans}
\end{table}

\section{Results - Photospheric Survey}
In this section we present the results of our photospheric survey. Using 
the lines listed in Table \ref{table:trans}, we measured a photospheric 
velocity of $81.5 \pm 1.9$, $37.4 \pm 1.8$, and 
$30.0 \pm 1.7$ km s$^{-1}$ for WD0455-282, WD0621-376, and WD2211-495 
respectively. In Table \ref{table:starparam} we summarise the measured 
stellar parameters for each star, including $T_{\mathrm{eff}}$, log $g$, 
$M_{*}$,$R_{*}$, and the photospheric abundances. The abundances are 
expressed as number fractions of hydrogen. All uncertainties quoted are 
to one sigma. In this section we first discuss the abundance 
measurements made for each species detected in the three white dwarfs. 
We then present measurements for $T_{\mathrm{eff}}$/log $g$ for each 
star. In conjunction with data from the astrometric satellite Gaia, we 
then give direct measurements of the distances, masses, radii, and 
gravitational redshifts for each object. 

\begin{table*}
\renewcommand{\arraystretch}{1.2}
\centering
\caption{List of measured parameters for WD0455-282, WD0621-376, and 
WD2211-495. RA, Dec, $G$, $G_{\mathrm{BP}}$, $G_{\mathrm{RP}}$, and 
Parallax data taken from Gaia DR2. Note that the RA/Dec coordinates are 
in the ICRS reference, and the epoch is J2015.5. Bolometric magnitudes,
and cooling ages were determined using the Montreal Cooling tables (see
\protect\citealt{holberg2006a}). }
\begin{tabular}[H]{@{}lccc}
\hline
Parameter                & WD0455-282                            & WD0621-376                            & WD2211-495 \\
\hline
$T_{\mathrm{eff}}$ (K)  & $65400_{-1210}^{+1240}$   & $64500_{-424}^{+437}$    & $65900_{-254}^{+259}$  \\
Log $g$ (cgs)           & $7.87_{-0.09}^{+0.09}$    & $7.60_{-0.03}^{+0.03}$   & $7.58_{-0.02}^{+0.02}$ \\
$v_{\mathrm{phot}}$ (km s$^{-1}$) & $81.5 \pm 1.9$  & $37.4 \pm 1.8$           & $30.0 \pm 1.7$         \\
RA (deg)                & $ 74.30575 \pm 0.25388$   & $ 95.80294 \pm 0.04720$  & $333.54973 \pm 0.04232$ \\
Dec (deg)               & $-28.13127 \pm 0.31315$   & $-37.69113 \pm 0.05733$  & $-49.32451 \pm 0.06189$ \\
Gaia $G$                & $13.915 \pm 0.002$       & $12.041 \pm 0.002$        & $11.610 \pm 0.002$     \\
Gaia $G_{\mathrm{BP}}$  & $13.651 \pm 0.009$       & $11.789 \pm 0.013$        & $11.315 \pm 0.007$     \\
Gaia $G_{\mathrm{RP}}$  & $14.236 \pm 0.003$       & $12.382 \pm 0.003$        & $11.951 \pm 0.002$     \\
Johnson $V$             & $13.901 \pm 0.023^{\dag}$   & $12.063 \pm 0.020^{\dag}$ & $11.667 \pm 0.031^{\dag}$ \\
$M_{\mathrm{Bol}}$      & $3.220 \pm 0.190$         & $2.778 \pm 0.064$        & $2.641 \pm 0.042$      \\
Cooling Age (Myr)       & $0.96 \pm 0.07$           & $0.82 \pm 0.03$          & $0.76 \pm 0.02$        \\
Parallax (mas)          & $8.237 \pm 0.354$         & $13.136 \pm 0.059$       & $16.992 \pm 0.082$     \\
Distance (pc)           & $121 \pm 5$               & $76.1 \pm 0.3$           & $58.9 \pm 0.3$         \\
Mass ($M_\odot$)        & $0.589 \pm 0.145$         & $0.731 \pm 0.053$        & $0.575 \pm 0.028$      \\
Radius (${R}_\odot/100$) & $1.48 \pm 0.06$          & $2.24 \pm 0.01$          & $2.04 \pm 0.01$        \\
$V_{\mathrm{grav}}$ (km s$^{-1}$) & $25.4 \pm 6.3$  & $20.7 \pm 1.5$           & $18.0 \pm 0.9$         \\
N(C)/N(H)               & $1.00\times{10}^{-8}$                 & $1.49_{-0.08}^{+0.08}\times{10}^{-6}$ & $5.05_{-0.55}^{+0.55}\times{10}^{-7}$ \\
N(N)/N(H)               & $7.71_{-0.21}^{+0.21}\times{10}^{-8}$ & $6.45_{-0.12}^{+0.12}\times{10}^{-7}$ & $1.49_{-0.05}^{+0.05}\times{10}^{-6}$ \\
N(O)/N(H)               & $1.58_{-0.12}^{+0.12}\times{10}^{-7}$ & $2.46_{-0.03}^{+0.03}\times{10}^{-6}$ & $2.00_{-0.03}^{+0.03}\times{10}^{-6}$ \\
N(Si)/N(H)              & $9.18_{-0.29}^{+0.29}\times{10}^{-7}$ & $8.75_{-0.19}^{+0.18}\times{10}^{-6}$ & $4.38_{-0.15}^{+0.15}\times{10}^{-6}$ \\
N(P)/N(H)               & $7.58_{-0.46}^{+0.46}\times{10}^{-8}$ & $1.70_{-0.16}^{+0.16}\times{10}^{-7}$ & $1.73_{-0.07}^{+0.07}\times{10}^{-7}$ \\
N(S)/N(H)               & $2.70_{-0.25}^{+0.25}\times{10}^{-7}$ & $1.05_{-0.06}^{+0.15}\times{10}^{-6}$ & $8.06_{-0.47}^{+0.46}\times{10}^{-7}$ \\
N(Fe)/N(H)              & $2.72_{-0.11}^{+0.11}\times{10}^{-6}$ & $1.84_{-0.04}^{+0.04}\times{10}^{-5}$ & $1.21_{-0.04}^{+0.04}\times{10}^{-5}$ \\
N(Ni)/N(H)              & $7.96_{-0.27}^{+0.27}\times{10}^{-7}$ & $4.78_{-0.08}^{+0.08}\times{10}^{-6}$ & $3.32_{-0.10}^{+0.10}\times{10}^{-6}$ \\
\hline
\multicolumn{4}{l}{$^\dag$The AAVSO Photometric All Sky Survey (APASS) DR9, \cite{henden2016a}.}\\
\end{tabular}
\label{table:starparam}
\renewcommand{\arraystretch}{1.0}
\end{table*}

\subsection{C Abundance}
\begin{figure}
\begin{centering}
\includegraphics[width=80mm]{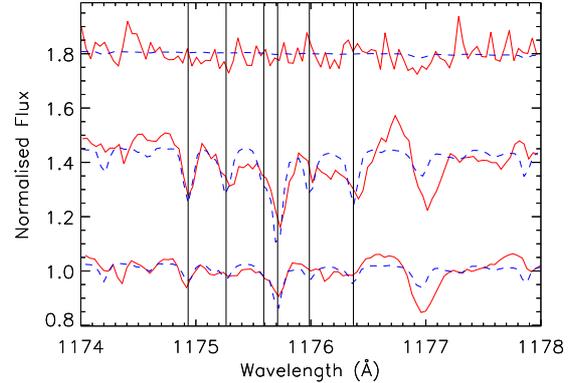}
\caption{Section of FUSE spectra for WD0455-282 (top), WD0621-376 
(middle), and WD2211-495 (bottom) containing the C {\sc iii} multiplet 
shifted to rest wavelength. The flux for each white dwarf has been 
normalised, and offset for clarity. The red solid line is the observed 
flux, while the blue dashed line is the best fitting photospheric model. 
The vertical solid lines indicate the positions of the transitions used 
to measure the abundance.}
\label{fig:threewd_ciii}
\end{centering}
\end{figure}

In Figure \ref{fig:threewd_ciii} we have plotted the FUSE spectra of the 
three white dwarfs spanning 1174-1178\AA\, which contains the 
C {\sc iii} multiplet transition. Interestingly, we find no traces of 
this multiplet in the spectrum of WD0455-282. Therefore, we place a limit 
on the photospheric C abundance in this star of $<1.00\times{10}^{-8}$. 
The C {\sc iii} multiplet is only just visible in the cases 
of WD0621-376 and WD2211-495. As mentioned previously, the uncertainties 
for the FUSE data are unreliable. Therefore, when measuring the C abundance, we 
expressed the flux uncertainty as a fraction of the observed flux so as 
to achieve a fit with $\chi_{\mathrm{red}}^{2}$ close to unity. We 
measure photospheric C abundances of 
$1.49_{-0.08}^{+0.08}\times{10}^{-6}$ and 
$5.05_{-0.55}^{+0.55}\times{10}^{-7}$ for WD0621-376 and WD2211-495 
respectively. 

\subsection{N Abundance}
\begin{figure}
\begin{centering}
\includegraphics[width=80mm]{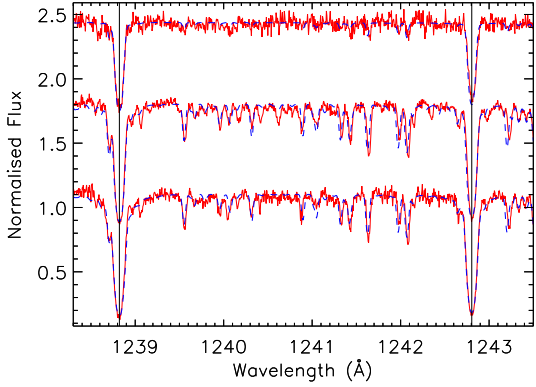}
\caption{Section of the STIS spectrum from WD0455-282 (top), WD0621-376 
(middle) and WD2211-495 (bottom) containing the N {\sc v} doublet. The 
fluxes have been normalised, and offset from each other for clarity. 
Note that the observed flux has been shifted to rest wavelength. The 
red-solid line is the observational data, while the blue dashed line is 
the best fitting photospheric model. The vertical black lines indicate 
the position of the N {\sc v} doublet used to measure the abundances.}
\label{fig:threewd_nv}
\end{centering}
\end{figure}

In Figure \ref{fig:threewd_nv} we have plotted a region of the STIS 
spectra for the three white dwarfs containing the N {\sc v} doublet. We 
detected the N {\sc iv} multiplet in the FUSE spectra at 914-918\AA\, 
for all three stars, however, the signal-to-noise in this region was too 
low to measure an accurate abundance. The N {\sc v} doublet is clearly 
resolved in all three stars, and we measure N abundances of 
$7.71_{-0.21}^{+0.21}\times{10}^{-8}$, 
$6.45_{-0.12}^{+0.12}\times{10}^{-7}$, and 
$1.49_{-0.05}^{+0.05}\times{10}^{-6}$ for WD0455-282, WD0621-376, and 
WD2211-495 respectively.

\subsection{O Abundance}
\begin{figure}
\begin{centering}
\includegraphics[width=80mm]{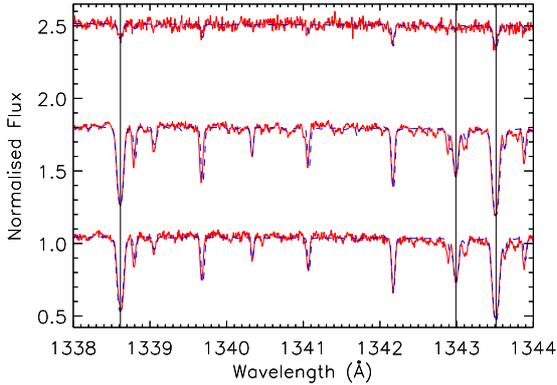}
\caption{The same as for Figure \protect\ref{fig:threewd_nv}, but for a 
region of the STIS spectrum containing three O {\sc iv} lines with 
laboratory wavelengths 1338.615, 1342.990, and 1343.514\AA\, 
respectively.}
\label{fig:threewd_oiv}
\end{centering}
\end{figure}

\begin{figure}
\begin{centering}
\includegraphics[width=80mm]{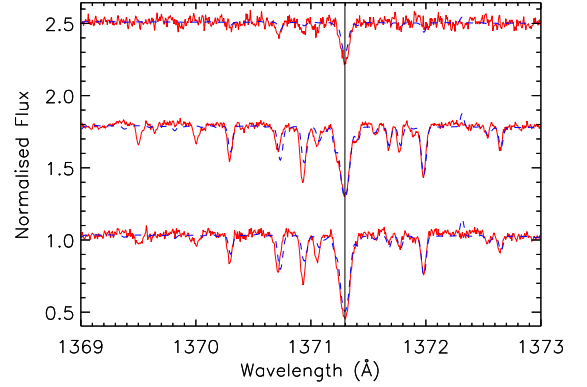}
\caption{The same as for Figure \protect\ref{fig:threewd_nv}, 
but for a region of the STIS spectrum containing an O {\sc v} line with 
laboratory wavelength 1371.296\AA.}
\label{fig:threewd_ov}
\end{centering}
\end{figure}

In Figure \ref{fig:threewd_oiv} we have plotted the region of STIS 
spectrum containing two of the O {\sc iv} transitions used to measure
the O abundance for WD0455-282, WD0621-376, and WD2211-495. The 
O {\sc v} line at 1371\AA\, was also detected, however, we opted to use 
the O {\sc iv} lines as there were three lines to fit. We measured O 
abundances of $1.58_{-0.12}^{+0.12}\times{10}^{-7}$, 
$2.46_{-0.03}^{+0.03}\times{10}^{-6}$, and 
$2.00_{-0.03}^{+0.03}\times{10}^{-6}$ for WD0455-282, WD0621-376, and 
WD2211-495 respectively. In Figure \ref{fig:threewd_ov} we 
have plotted a region of STIS spectrum containing the O {\sc v} 
1371.296\AA\, line for the three white dwarfs. Excellent agreement is 
seen between the synthetic and observed spectrum for WD0621-376. 
However, for WD0455-282 and WD2211-495 it can be seen that the synthetic 
profile for O {\sc v} doesn't descend all the way into the observed profile. 
This implies the $T_{\mathrm{eff}}$ may be underestimated for these 
stars. This may be due to our convergence criteria being too relaxed, 
where iterations of $T_{\mathrm{eff}}$, log $g$, and abundance 
measurements were performed until the difference between the previous 
and current measurements was less than the uncertainty on the 
measurement.

\subsection{Si Abundance}\label{siabunsec}
\begin{figure}
\begin{centering}
\includegraphics[width=80mm]{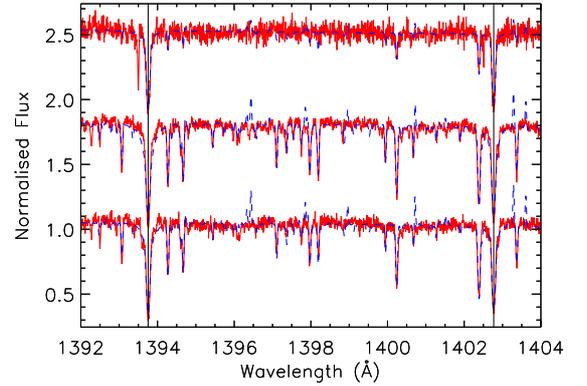}
\caption{The same as for Figure \protect\ref{fig:threewd_nv}, but for a 
region of the STIS spectrum containing two Si {\sc iv} lines with 
laboratory wavelengths 1393.755 and 1402.770\AA. For WD0455-282, the 
additional absorption feature next to the photospheric profile is a 
circumstellar line, whose discussion is deferred to Section 
\ref{circumsec}.}
\label{fig:threewd_siiv}
\end{centering}
\end{figure}

In Figure \ref{fig:threewd_siiv} we have plotted the region of STIS 
spectrum containing the resonant Si {\sc iv} doublet for WD0455-282, 
WD0621-376, and WD2211-495. In addition to the STIS data, we also used 
two excited Si {\sc iv} absorption features from the FUSE spectrum. As 
was the case with the C {\sc iii} transition, we expressed the FUSE flux 
uncertainty as a fraction of the observed flux so as to achieve a fit 
with $\chi_{\mathrm{red}}^{2}$ close to unity. We measured Si abundances 
of $9.18_{-0.29}^{+0.29}\times{10}^{-7}$, 
$8.75_{-0.19}^{+0.18}\times{10}^{-6}$, and
$4.38_{-0.15}^{+0.15}\times{10}^{-6}$ for WD0455-282, WD0621-376, and 
WD2211-495 respectively. We note that our Si abundance 
measurements for WD0621-376 and WD2211-495 are an order of magnitude 
larger than that of B14, who measured 
$7.57_{-0.31}^{+0.31}\times{10}^{-7}$ and 
$2.41_{-2.04}^{+5.21}\times{10}^{-7}$ respectively. For WD0621-376 we 
defer discussion until later in the paper. However, in the case of 
WD2211-495 this large difference may be due to the higher quality 
observational data used in this work. For comparison, B14 used a FUSE
spectrum constructed with one dataset (M10303160), whereas the present
work uses a FUSE spectrum with 18 datasets (see Table 
\ref{table:fuseobs}).

\subsection{P Abundance}
\begin{figure}
\begin{centering}
\includegraphics[width=80mm]{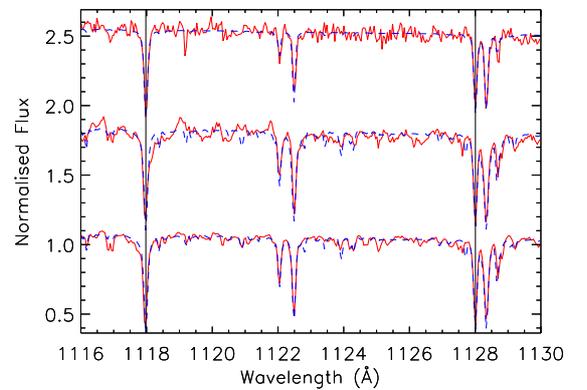}
\caption{The same as for Figure \protect\ref{fig:threewd_ciii}, but for a region of the FUSE spectrum
containing two P {\sc v} lines with laboratory wavelengths 1117.977 and 1128.008\AA\, respectively.}
\label{fig:threewd_pv}
\end{centering}
\end{figure}

In Figure \ref{fig:threewd_pv} we have plotted the region of FUSE 
spectra for the three white dwarfs that contains the P {\sc v} resonant 
doublet. As was the case for the C {\sc iii} multiplet, we expressed the 
flux uncertainties in the FUSE data as a fraction of the observed flux 
so as to achieve a fit with $\chi_{\mathrm{red}}^{2}$ close to unity. We 
measured the photospheric P abundance to be 
$7.58_{-0.46}^{+0.46}\times{10}^{-8}$, 
$1.70_{-0.16}^{+0.16}\times{10}^{-7}$, and 
$1.73_{-0.07}^{+0.07}\times{10}^{-7}$ for WD0455-282, WD0621-376, and 
WD2211-495 respectively.

\subsection{S Abundance}
\begin{figure}
\begin{centering}
\includegraphics[width=80mm]{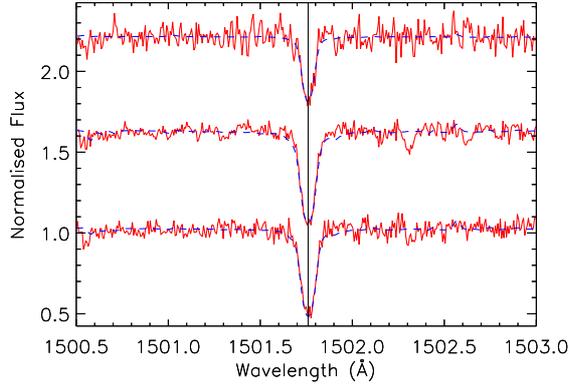}
\caption{The same as for Figure \protect\ref{fig:threewd_nv}, but for a region of the STIS spectrum 
containing the S {\sc v} line with laboratory wavelength 1501.760\AA.}
\label{fig:threewd_sv}
\end{centering}
\end{figure}

In Figure \ref{fig:threewd_sv} we have plotted the region of STIS spectrum containing the excited
S {\sc v} transition for the three white dwarfs. The resonant S {\sc vi} doublet is detected in the FUSE
spectra, however, we opted not to use these for abundance measurements. This is because they are
situated close to the line centroids of the Lyman series where the wings are changing rapidly, meaning
the abundance will be sensitive to the broadening of the Lyman lines. We measured an S abundance 
of $2.70_{-0.25}^{+0.25}\times{10}^{-7}$, $1.05_{-0.06}^{+0.15}\times{10}^{-6}$, and 
$8.06_{-0.47}^{+0.46}\times{10}^{-7}$ for WD0455-282, WD0621-376, and WD2211-495 respectively.

\subsection{Fe Abundance}
\begin{figure}
\begin{centering}
\includegraphics[width=80mm]{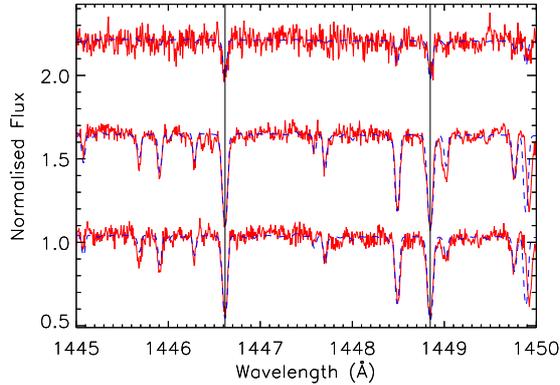}
\caption{The same as for Figure \protect\ref{fig:threewd_nv}, but for a
region of STIS spectrum containing two Fe {\sc v} lines with laboratory 
wavelengths 1446.617 and 1448.847\AA\, respectively.}
\label{fig:threewd_fev}
\end{centering}
\end{figure}

\begin{figure}
\begin{centering}
\includegraphics[width=80mm]{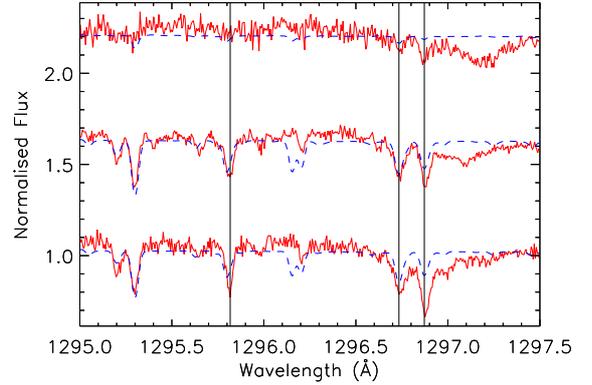}
\caption{The same as for Figure \protect\ref{fig:threewd_nv}, but for a
region of STIS spectrum containing three Fe {\sc vi} lines with laboratory 
wavelengths 1295.817, 1296.734 and 1296.872\AA\, respectively.}
\label{fig:threewd_fevi}
\end{centering}
\end{figure}

In Figure \ref{fig:threewd_fev} we have plotted a region of the STIS 
spectrum containing two of the Fe {\sc v} lines used to measure the Fe 
abundance in the three white dwarfs. We measured Fe abundances of 
$2.72_{-0.11}^{+0.11}\times{10}^{-6}$, 
$1.84_{-0.04}^{+0.04}\times{10}^{-5}$, and
$1.21_{-0.04}^{+0.04}\times{10}^{-5}$ for WD0455-282, WD0621-376, and 
WD2211-495 respectively. In Figure \ref{fig:threewd_fevi}
we have also plotted a region of STIS spectrum containing three Fe 
{\sc vi} absorption features with laboratory wavelengths 1295.817, 
1296.734 and 1296.872\AA\, respectively for the three white dwarfs. As 
with the case of O {\sc v}, excellent agreement is seen between the 
observed and synthetic spectra for WD0621-376, while the synthetic 
spectra for WD0455-282 and WD2211-495 do not descend completely into the
observed spectrum.

\subsection{Ni Abundance}
\begin{figure}
\begin{centering}
\includegraphics[width=80mm]{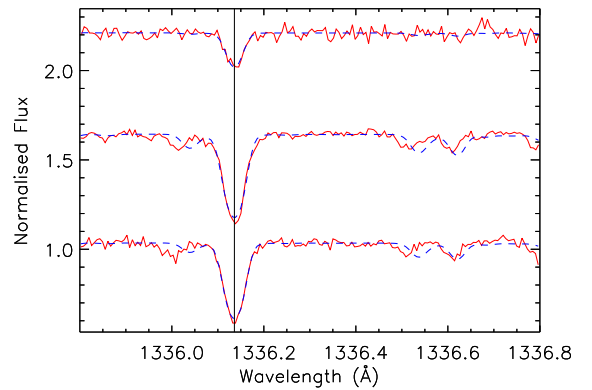}
\caption{The same as for Figure \protect\ref{fig:threewd_nv}, but for a 
region of STIS spectrum containing an Ni {\sc v} line with laboratory 
wavelength 1336.136\AA.}
\label{fig:threewd_niv}
\end{centering}
\end{figure}

In Figure \ref{fig:threewd_niv} we have plotted a region of the STIS 
spectrum containing one of the Ni {\sc v} features used to measure the 
Ni abundance in the three white dwarfs. We measured Ni abundances
of $7.96_{-0.27}^{+0.27}\times{10}^{-7}$, 
$4.78_{-0.08}^{+0.08}\times{10}^{-6}$, and 
$3.32_{-0.10}^{+0.10}\times{10}^{-6}$ for WD0455-282, WD0621-376, and 
WD2211-495 respectively.

\subsection{$T_{\mathrm{eff}}$/log $g$ measurements}\label{lymsec}
The Lyman lines (Ly-$\beta$ - Ly-$\epsilon$) used to measure 
$T_{\mathrm{eff}}$ and log $g$ are located in the FUSE spectrum. As 
mentioned previously, the flux uncertainties in the FUSE spectra are 
unreliable. Therefore, we express the flux uncertainties as a fraction 
of the observational flux so as to achieve a fit with 
$\chi_{\mathrm{red}}^{2}$ close to unity. For $T_{\mathrm{eff}}$, we 
measured values of $65400_{-1210}^{+1240}$, $64500_{-424}^{+437}$, and 
$65900_{-254}^{+259}$K for WD0455-282, WD0621-376, and WD2211-495 
respectively. For log $g$, we measured values of $7.87_{-0.09}^{+0.09}$, 
$7.60_{-0.03}^{+0.03}$, and $7.58_{-0.02}^{+0.02}$ for WD0455-282, 
WD0621-376, and WD2211-495 respectively. Interestingly, we noted that 
the log $g$ measurement for WD0621-376 is much higher than reported in 
previous works. Indeed, B14 measured log $g=7.22$ for this star. 

\subsection{Mass-radius measurements}
Prior to the launch of the astrometric satellite Gaia, measurement of 
astrophysical parameters such as the mass and radius of a white dwarf 
required the use of theoretical cooling tables such as those calculated 
by \citet{holberg2006a,kowalski2006a,tremblay2011a,bergeron2011a} 
\footnote{http://www.astro.umontreal.ca/{\textasciitilde}bergeron/CoolingModels}
(Montreal cooling tables hereafter). Thanks to the highly accurate 
parallax measurements made by Gaia, the angular diameter of any field 
white dwarf can be measured to a high degree of precision. Combined with 
the distance, we can then extract the radius, and in conjunction with 
log $g$, we can then obtain the mass, followed by the gravitational 
redshift velocity. These mass and radius measurements offer a valuable 
means with which to test the fundamental mass-radius relationship for 
white dwarf stars. 

For WD0455-282, the Gaia DR2 parallax was measured to be 
$8.237 \pm 0.354$mas, corresponding to a distance of $121 \pm 5$pc. 
Using equations \ref{eqn:rad} and \ref{eqn:mass}, we extracted a mass 
and radius of $0.589 \pm 0.145M_{\odot}$ and 
$1.48 \pm 0.06R_{\odot}/100$ respectively. The large uncertainty on 
the mass is dominated by the uncertainty on our log $g$ measurement. 
With the mass and radius, we then calculated a gravitational redshift 
velocity of $25.4 \pm 6.3$ km s$^{-1}$.

For WD0621-376, the Gaia DR2 parallax is $13.136 \pm 0.059$mas, 
corresponding to a distance of $76.1 \pm 0.3$pc. With these values, we 
measured a mass and radius of $0.731 \pm 0.053M_\odot$ and 
$2.24 \pm 0.01R_{\odot}/100$ respectively. Using the mass and radius, 
we measured the gravitational redshift velocity to be 
$20.7 \pm 1.5$ km s$^{-1}$. We are skeptical of our mass measurement due 
to the large difference between our measured log $g$ (7.60), and that of 
\cite{barstow2014a} (7.22). If we use log $g$=7.22 in Equation 
\ref{eqn:mass} then the measured mass decreases to $0.305M_{\odot}$. We 
defer further exploration of this issue to the discussion.

For WD2211-495 the measured Gaia DR2 parallax is $16.992 \pm 0.082$mas, 
corresponding to a distance of $58.9 \pm 0.3$pc, making the star the 
closest in our sample. With this information, we measured the mass and 
radius to be $0.575 \pm 0.028M_{\odot}$ and 
$2.04 \pm 0.01R_{\odot}/100$ respectively. Using the mass and radius, 
we then calculated the gravitational redshift velocity to be 
$18.0 \pm 0.9$ km s$^{-1}$.

\section{Results - Interstellar Survey}\label{ismsec}
In this section we present the results of our interstellar survey. To 
aid classification of the detected features, we used the Dynamical Local 
ISM calculator, the theory and motivation for which is described by 
\cite{redfield2008a}. The calculator is freely available 
online\footnote{http://lism.wesleyan.edu/LISMdynamics.html}, and gives 
an indication of what interstellar clouds may be traversed along the 
line of sight given a galactic latitude, and longitude. The calculator 
also predicts the velocity components of the cloud in question. In Table 
\ref{table:ismparam} we summarise the detected velocity components, and 
the clouds intersected by the line of sight.

\begin{table*}
\renewcommand{\arraystretch}{1.2}
\centering
\caption{List of measured ISM velocities for WD0455-282, WD0621-376, and 
WD2211-495, and the (known) clouds intersected by the line of sight.}
\begin{tabular}[H]{@{}lccc}
\hline
Parameter              & WD0455-282                         & WD0621-376                         & WD2211-495                           \\
\hline
No of components       & 4                                  & 2                                  & 2                                    \\
$V_{i}$ (km s$^{-1}$)  & $-40.3 \pm 1.6$, $-32.1 \pm 3.0$ & $ 12.1 \pm 2.3$, $ 23.1 \pm 2.4$ & $ -9.91 \pm 1.39$, $ -2.18 \pm 2.43$ \\
                       & $ 16.7 \pm 1.9$, $ 25.1 \pm 1.1$ &                                    &                                      \\
Clouds intersected     & Blue                               & Blue                               & LIC                                  \\
\hline
\end{tabular}
\label{table:ismparam}
\renewcommand{\arraystretch}{1.0}
\end{table*}

\subsection{WD0455-282}
\begin{figure}
\begin{centering}
\includegraphics[width=80mm]{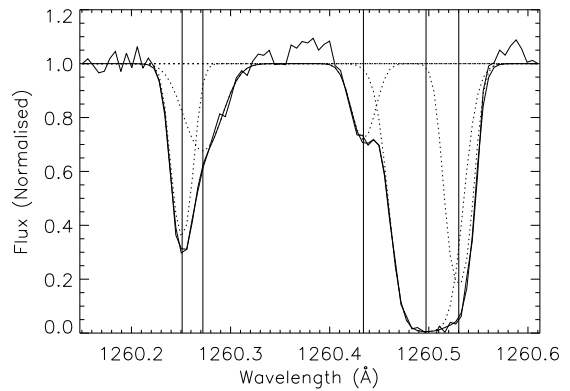}
\caption{Best fit to the absorption features detected between 1260.2 to 
1260.7\AA\, using 6 Gauss profiles for WD0455-282. From left to right, 
the first two features correspond to interstellar Si {\sc ii} with 
velocities of $-40.7 \pm 0.2$ and $-35.7 \pm 2.2$ km s$^{-1}$ 
respectively. The third feature is a photospheric line, and the last 
two features are interstellar Si {\sc ii} with velocities of 
$17.9 \pm 1.2$ and $25.7 \pm 1.3$ km s$^{-1}$ respectively.}
\label{fig:wd0455si2}
\end{centering}
\end{figure}

The line of sight to WD0455-282 appears to have a very interesting 
structure. In our line survey we detected four distinct ISM velocity 
components, with average velocities of $v_{1}=-40.3 \pm 1.6$, 
$v_{2}=-32.1 \pm 3.0$, $v_{3}=16.7 \pm 1.9$, and 
$v_{4}=25.1 \pm 1.1$ km s$^{-1}$ respectively. In Figure 
\ref{fig:wd0455si2} we have plotted a region of STIS spectrum for 
WD0455-282 containing the Si {\sc ii} 1260\AA\, transition, where all 
four velocity components can be seen. We measured the four velocities to 
be $-40.7 \pm 0.2$, $-35.7 \pm 2.2$, $17.9 \pm 1.2$, and 
$25.7 \pm 1.3$ km s$^{-1}$ respectively. The LISM calculator states that 
the light of sight towards WD0455-282 traverses the Blue cloud, which 
has a radial velocity of $12.6 \pm 1.0$ km s$^{-1}$. While this 
velocity is close to our measurement for $v_{3}$, the two values are not
in agreement.

\begin{figure}
\begin{centering}
\includegraphics[width=80mm]{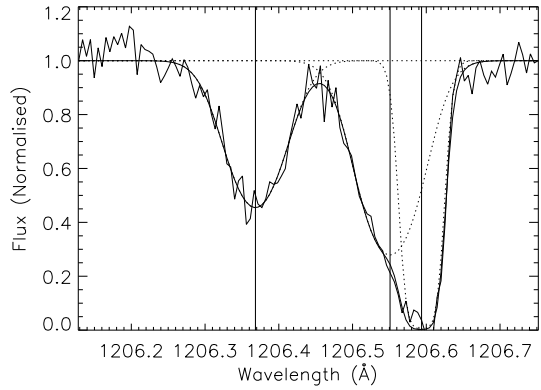}
\caption{Best fit to the absorption features detected 
between 1206.2 and 1206.7\AA\, using 3 Gauss profiles for WD0455-282. 
All three features arise from Si {\sc iii} 1206\AA, with velocities of 
$-32.8 \pm 0.4$, $12.8 \pm 1.3$, and $23.4 \pm 0.3$ km s$^{-1}$. All
three components are circumstellar.}
\label{fig:wd0455si3ism1}
\end{centering}
\end{figure}

\begin{figure}
\begin{centering}
\includegraphics[width=80mm]{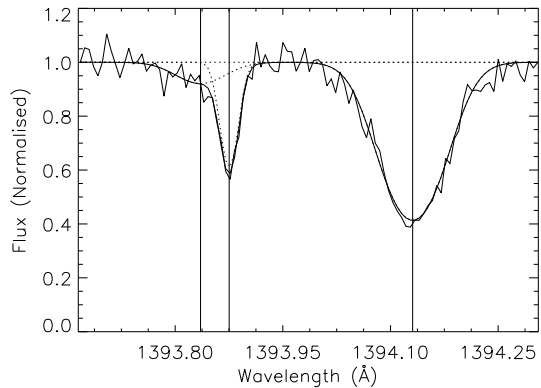}
\caption{Best fit to the absorption features detected 
between 1393.7 and 1394.3\AA\, using 3 Gauss profiles for WD0455-282. 
All three features arise from Si {\sc iv} 1393\AA, with velocities 
$17.3 \pm 4.9$, $25.9 \pm 0.2$, and $80.9 \pm 0.2$ km s$^{-1}$. The 
first two components are interstellar, and the third component is 
photospheric.}
\label{fig:wd0455si4ism1}
\end{centering}
\end{figure}

\begin{figure}
\begin{centering}
\includegraphics[width=80mm]{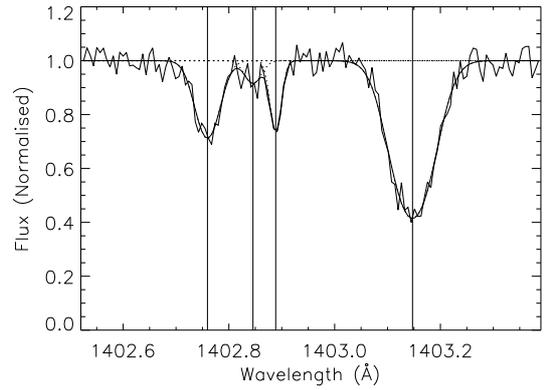}
\caption{Best fit to the absorption features detected 
between 1402.5 and 1403.4\AA\, using 3 Gauss profiles for WD0455-282. 
The latter three features arise from Si {\sc iv} 1402\AA, with 
velocities of $16.1 \pm 1.4$, $25.3 \pm 0.4$, and 
$80.6 \pm 0.2$ km s$^{-1}$ respectively. The absorption feature on the 
far left is a photospheric Fe line. The second and third components are 
circumstellar Si {\sc iv}, and the fourth component is photospheric 
Si {\sc iv}.}
\label{fig:wd0455si4ism2}
\end{centering}
\end{figure}

Interestingly, our survey of WD0455-282 has revealed the presence of 
multi-component circumstellar absorption in the Si {\sc iii} 1206, 
Si {\sc iv} 1393, and 1402\AA\, resonant lines. In Figs 
\ref{fig:wd0455si3ism1}, \ref{fig:wd0455si4ism1}, and 
\ref{fig:wd0455si4ism2} we have plotted three spectral regions 
containing the Si {\sc iii} and Si {\sc iv} resonant lines, along with 
our best fit to these data using Gaussian profiles. The Si {\sc iii} 
1206\AA\, line has three circumstellar components with velocities of
$-32.8 \pm 0.4$, $12.8 \pm 1.3$, and $23.4 \pm 0.3$ km s$^{-1}$
respectively. Si {\sc iv} 1393\AA\, has two circumstellar components 
with velocities of $17.3 \pm 4.9$ and $25.9 \pm 0.2$ respectively.
Lastly, the Si {\sc iv} 1402\AA\, line has two circumstellar components 
with velocities of $16.1 \pm 1.4$ and $25.3 \pm 0.4$ respectively.
For the Si {\sc iii} line, the circumstellar velocities are strikingly 
similar to those measured for the ISM velocities $v_{2}$, $v_{3}$, and 
$v_{4}$. For Si {\sc iv} the circumstellar velocities are similar to 
those measured for $v_{3}$ and $v_{4}$. To our knowledge, this is the 
first time that multi-component circumstellar absorption has been 
observed along the line of sight to a white dwarf star. 

\subsection{WD0621-376}
\begin{figure}
\begin{centering}
\includegraphics[width=80mm]{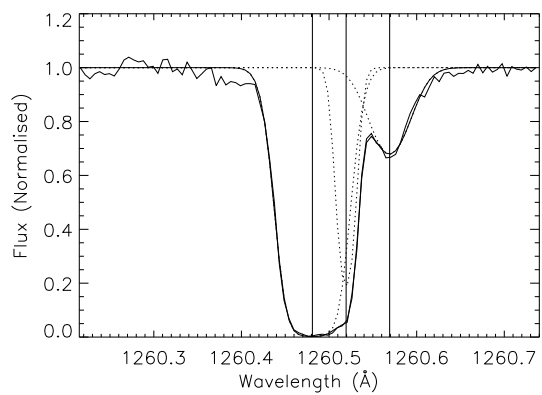}
\caption{Best fit to the absorption features detected 
between 1260.2 to 1260.7\AA\, using three Gauss profiles for WD0621-376.
The first two features are interstellar Si {\sc ii} with velocities of 
$14.0 \pm 0.3$ and $23.2 \pm 0.3$ km s$^{-1}$ respectively, while the 
third feature is a photospheric line.}
\label{fig:wd0621si2}
\end{centering}
\end{figure}

We made detections of two distinct ISM velocity components along the 
line of sight to WD0621-376 with average velocities of $v_{1}=12.1 \pm 2.3$, 
and $v_{2}=23.1 \pm 2.4$ km s$^{-1}$ respectively. The LISM calculator 
indicated the line of sight to WD0621-376 intersects the Blue cloud, 
with a predicted radial velocity of $11.1 \pm 0.9$ km s$^{-1}$. Based on 
this value, our first ISM component appears to originate from the Blue 
cloud. In Figure \ref{fig:wd0621si2} we have plotted a region of 
spectrum containing the Si {\sc ii} 1260\AA\, transition, where both ISM 
components can be seen with velocities of $14.0 \pm 0.3$ and 
$23.2 \pm 0.3$ km s$^{-1}$ respectively.

Unlike WD0455-282, we didn't detect any circumstellar features along the 
line of sight to WD0621-376. In addition, we made no detections of the 
Si {\sc iii} 1206.4995\AA\, transition. The significance of this result 
will be explored shortly.

\subsection{WD2211-495}
\begin{figure}
\begin{centering}
\includegraphics[width=80mm]{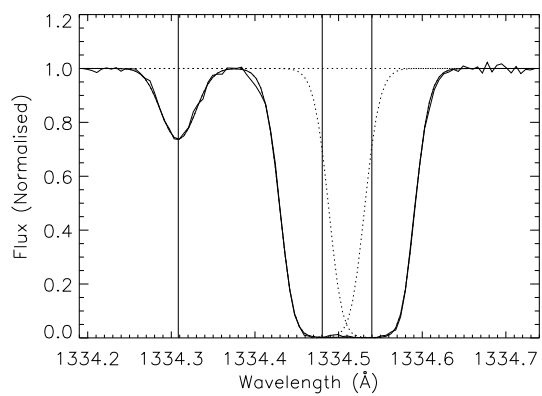}
\caption{Best fit to the absorption features detected between 1334.2 to 
1334.7\AA\, using three Gauss profiles for WD2211-495. The first feature
is photospheric, while the second and third features are interstellar 
C {\sc ii} with velocities of $-11.7 \pm 0.8$, and 
$1.67 \pm 0.85$ km s$^{-1}$ respectively.}
\label{fig:wd2211c2}
\end{centering}
\end{figure}

We have identified two distinct ISM velocity components along the line 
of sight to WD2211-495, with average velocities of 
$v_{1}=-9.91 \pm 1.39$ and $v_{2}=-2.18 \pm 2.43$ km s$^{-1}$ 
respectively. In Figure \ref{fig:wd2211c2} we have plotted a region of 
STIS spectrum for WD2211-495 containing the C {\sc ii} 1334\AA\, line, 
where both ISM components can be seen with velocities of $-11.7 \pm 0.8$ 
and $1.67 \pm 0.85$ km s$^{-1}$ respectively. The LISM calculator 
indicated that the line of sight traversed two clouds, namely the Local 
Interstellar Cloud (LIC), and the Dor cloud, with velocities $-8.80 \pm 1.32$ and 
$9.93 \pm 0.60$ km s$^{-1}$ respectively. Our first ISM velocity is in 
agreement with the value predicted for the LIC, suggesting this 
component is associated with the LIC. However, our last component is not 
in agreement with the Dor cloud velocity. This maybe due to a limitation 
in the LISM model, which is based upon four sightlines for the Dor Cloud 
(contrast with 79 sightlines for the LIC, and 10 for the Blue cloud). 
This could also be explained if the Dor cloud does not intersect the 
line of sight to the white dwarf. 

\begin{figure}
\begin{centering}
\includegraphics[width=80mm]{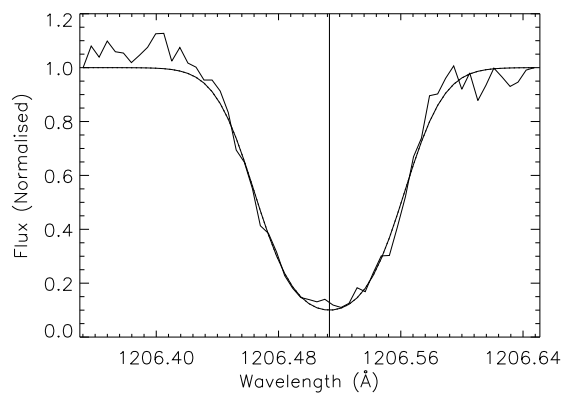}
\caption{Best fit to the absorption features detected 
between 1206 and 1207\AA\, using 1 Gauss profile for WD2211-495. This 
feature is circumstellar Si {\sc iii}, with a velocity of 
$3.43 \pm 0.10$ km s$^{-1}$.}
\label{fig:wd2211si3ism1}
\end{centering}
\end{figure}

\begin{figure}
\begin{centering}
\includegraphics[width=80mm]{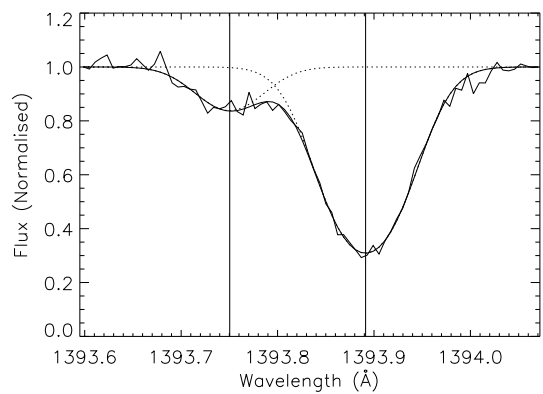}
\caption{Best fit to the absorption features detected 
between 1393.7 and 1394.3\AA\, using 2 Gauss profiles for WD2211-495.
The first feature is circumstellar Si {\sc iv} with a velocity of 
$-1.00 \pm 0.56$ km s$^{-1}$, while the second feature is photospheric
Si {\sc iv} with a velocity of $29.3 \pm 0.12$ km s$^{-1}$.}
\label{fig:wd2211si4ism1}
\end{centering}
\end{figure}

We made a detection of a single circumstellar component in the 
Si {\sc iii} 1206 line, and the Si {\sc iv} 1393\AA\, line. We do not 
detect a circumstellar line in the Si {\sc iv} 1402\AA\, line due to the 
presence of strong Fe/Ni absorption features. In Figs 
\ref{fig:wd2211si3ism1} and \ref{fig:wd2211si4ism1} we have plotted 
regions of STIS spectra for WD2211-495 encompassing the Si {\sc iii} and 
Si {\sc iv} transitions respectively. We measured the velocity of the 
circumstellar component to be $3.43 \pm 0.10$ km s$^{-1}$ for the 
Si {\sc iii} line, and $-1.00 \pm 0.56$ km s$^{-1}$ for the Si {\sc iv} 
line. The measured circumstellar velocities are similar to the velocity 
measured for the ISM component $v_{2}$.

\section{Discussion: Radiative levitation}\label{radlevsec}
\begin{figure}
\begin{centering}
\includegraphics[width=80mm]{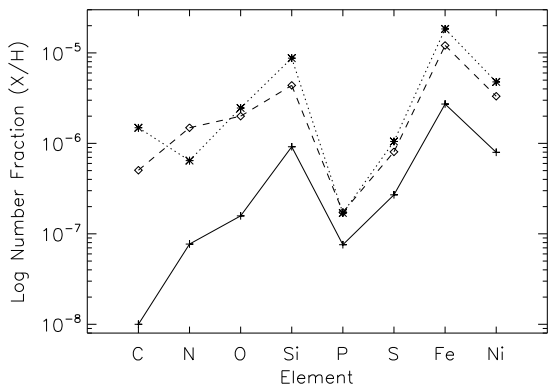}
\caption{Plot of observed photospheric abundances for WD0455-282 (solid 
line, plus-symbols), WD0621-376 (dotted line, asterisk-symbols), and 
WD2211-495 (dashed line, diamond-symbols). Note that the C abundance for 
WD0455-282is an upper limit.}
\label{fig:wdabn}
\end{centering}
\end{figure}

It is well established that metals are able to be supported against 
gravitational diffusion in white dwarf atmospheres thanks to radiative
levitation. The efficiency of radiative levitation is a sensitive 
function of $T_{\mathrm{eff}}$ and log $g$, meaning the observed 
abundance patterns in a white dwarf atmosphere will also be a function
of $T_{\mathrm{eff}}$ and log $g$ (neglecting accretion). In Figure 
\ref{fig:wdabn} we have plotted the measured abundances in all three 
white dwarfs as number fractions of H. A general trend can be seen for 
all three objects, where the abundance  of the metals increases steadily 
from C to Si, followed by a drop for the P abundance, followed by a 
steady increase again for S to Fe. It can also be seen that the 
abundance curves for WD0621-376 and WD2211-495 overlap one another. 
Furthermore, because WD0455-282 has a similar abundance pattern to the 
other two white dwarfs, it appears that the WD0455-282 curve can be 
brought up to meet the other two curves through the application of a 
multiplicative constant. This is simply explained by considering the 
$T_{\mathrm{eff}}$ and log $g$ values for each white dwarf. The measured
$T_{\mathrm{eff}}$ and log $g$ for WD0621-376 and WD2211-495 are very 
similar. Therefore, material in the atmospheres of these two stars will
experience a similar radiative acceleration. WD0455-282 has a similar 
$T_{\mathrm{eff}}$, but a higher log $g$, meaning the equilibrium
abundances will be smaller in this object compared to the other two. 
This point is further reinforced by considering the Fe/Ni abundance 
ratio. The Fe/Ni ratios observed in WD0455-282, WD0621-376, and 
WD2211-495 were calculated to be $3.42_{-0.18}^{+0.18}$, 
$3.85_{-0.10}^{+0.10}$, and $3.64_{-0.16}^{+0.16}$ respectively. Within
error the Fe/Ni ratio observed for WD2211-495 is in agreement with that
observed for WD0621-376, while the ratio observed for WD0621-376 is in 
agreement with that observed for WD0455-282.

There is continuing disagreement between the photospheric abundances 
observed in white dwarf atmospheres, and the photospheric abundances 
predicted by radiative levitation theory. The most well-known set of 
radiative equilibrium abundances for white dwarf stars were calculated 
by Ch94/95. This disagreement is thought to be due to a number of 
simplifying assumptions used in calculating the abundances. These 
assumptions include assuming no momentum redistribution of levitated 
material, and that the reservoir of material that could be levitated was 
effectively infinite. However, these values are useful to compare 
against, as they define the maximum possible abundance that can be 
supported in the atmosphere through radiative levitation. In 
Table \ref{table:radeq} we have tabulated the predicted abundances from 
radiative equilibrium theory, and compare them to the measured 
photospheric abundances from WD0455-282, WD0621-376, and WD2211-495. In 
all three stars, it can be seen that for C, N, O, P, S, and Ni the 
measured abundances are smaller than those predicted by Ch94/95. 
However, in the case of Si, the observed abundance is larger by several 
dex than those predicted by Ch94/95. As mentioned above, the abundances 
from radiative equilibrium theory will be the maximum possible abundance 
that can be supported through radiative levitation alone. Therefore, 
material exceeding this quantity should sink into the core. This implies 
that the reservoir of Si and Fe is being continually replenished.

In addition to the calculations performed by Ch94/95, more detailed 
models were considered by \cite{schuh2005a} that included the effects of 
gravitational diffusion and radiative levitation in a self consistent
manner. Schuh was able to show that the radiative support for Si 
plummets for DA white dwarfs with $T_{\mathrm{eff}}>60,000$K. This is 
because the population of Si {\sc iv} decreases as Si {\sc v} becomes 
populated with increasing temperature. This means there should be little 
to no Si present in the white dwarfs considered in this work. This 
corroborates the prediction by Ch94/95, providing further support
to the idea that the three objects are potentially accreting material.
However, it is not clear where this material originates from. The 
interplay between the flow of material sinking into the core via 
gravitational diffusion, the retardation of this flow by radiative 
levitation, and the introduction of additional material through 
accretion makes the problem complex. Therefore, this means 
that the methods used in studying metal pollution in cool white dwarfs 
(where radiative levitation is inefficient) cannot be applied to the 
stars studied in this paper. Future work will focus on developing the 
necessary computational and theoretical framework to study these 
processes, and how they affect the observed photospheric abundance 
patterns in white dwarf star atmospheres.

\begin{table*}
\renewcommand{\arraystretch}{1.2}
\centering
\caption{Comparison between predicted photospheric abundances from 
radiative levitation theory (Ch94/95), and those 
measured in this work (Observed column) for WD0455-282, WD0621-376, and
WD2211-495, as a number fraction relative to H.}
\begin{tabular}[H]{@{}lcccccc}
\hline
           & WD0455-282           &                      & WD0621-376           &                      & WD2211-495           &                      \\
N(X)/N(H)  & Ch94/95              & Observed             & Ch94/95              & Observed             & Ch94/95              & Observed             \\
\hline
 N(C)/N(H) & $1.51\times{10}^{-6}$ & $1.00\times{10}^{-8}$ & $2.82\times{10}^{-6}$ & $1.49\times{10}^{-6}$ & $3.02\times{10}^{-6}$ & $5.05\times{10}^{-7}$ \\
 N(N)/N(H) & $3.72\times{10}^{-6}$ & $7.71\times{10}^{-8}$ & $6.76\times{10}^{-6}$ & $6.45\times{10}^{-7}$ & $7.76\times{10}^{-6}$ & $1.49\times{10}^{-6}$ \\
 N(O)/N(H) & $4.07\times{10}^{-6}$ & $1.58\times{10}^{-7}$ & $7.24\times{10}^{-6}$ & $2.46\times{10}^{-6}$ & $8.32\times{10}^{-6}$ & $2.00\times{10}^{-6}$ \\
N(Si)/N(H) & $5.75\times{10}^{-9}$ & $9.18\times{10}^{-7}$ & $2.14\times{10}^{-8}$ & $8.75\times{10}^{-6}$ & $1.91\times{10}^{-8}$ & $4.38\times{10}^{-6}$ \\
 N(P)/N(H) & $1.78\times{10}^{-6}$ & $7.58\times{10}^{-8}$ & $3.72\times{10}^{-6}$ & $1.70\times{10}^{-7}$ & $4.07\times{10}^{-6}$ & $1.73\times{10}^{-7}$ \\
 N(S)/N(H) & $2.57\times{10}^{-6}$ & $2.70\times{10}^{-7}$ & $5.13\times{10}^{-6}$ & $1.05\times{10}^{-6}$ & $5.62\times{10}^{-6}$ & $8.06\times{10}^{-7}$ \\
N(Fe)/N(H) & $4.37\times{10}^{-6}$ & $2.72\times{10}^{-6}$ & $1.02\times{10}^{-5}$ & $1.84\times{10}^{-5}$ & $1.15\times{10}^{-5}$ & $1.21\times{10}^{-5}$ \\
N(Ni)/N(H) & $1.07\times{10}^{-4}$ & $7.96\times{10}^{-7}$ & $2.09\times{10}^{-4}$ & $4.78\times{10}^{-6}$ & $2.51\times{10}^{-4}$ & $3.32\times{10}^{-6}$ \\
\hline
\end{tabular}
\label{table:radeq}
\renewcommand{\arraystretch}{1.0}
\end{table*}

\section{Discussion: The mass-radius relationship}
\begin{figure*}
\begin{centering}
\includegraphics[width=180mm]{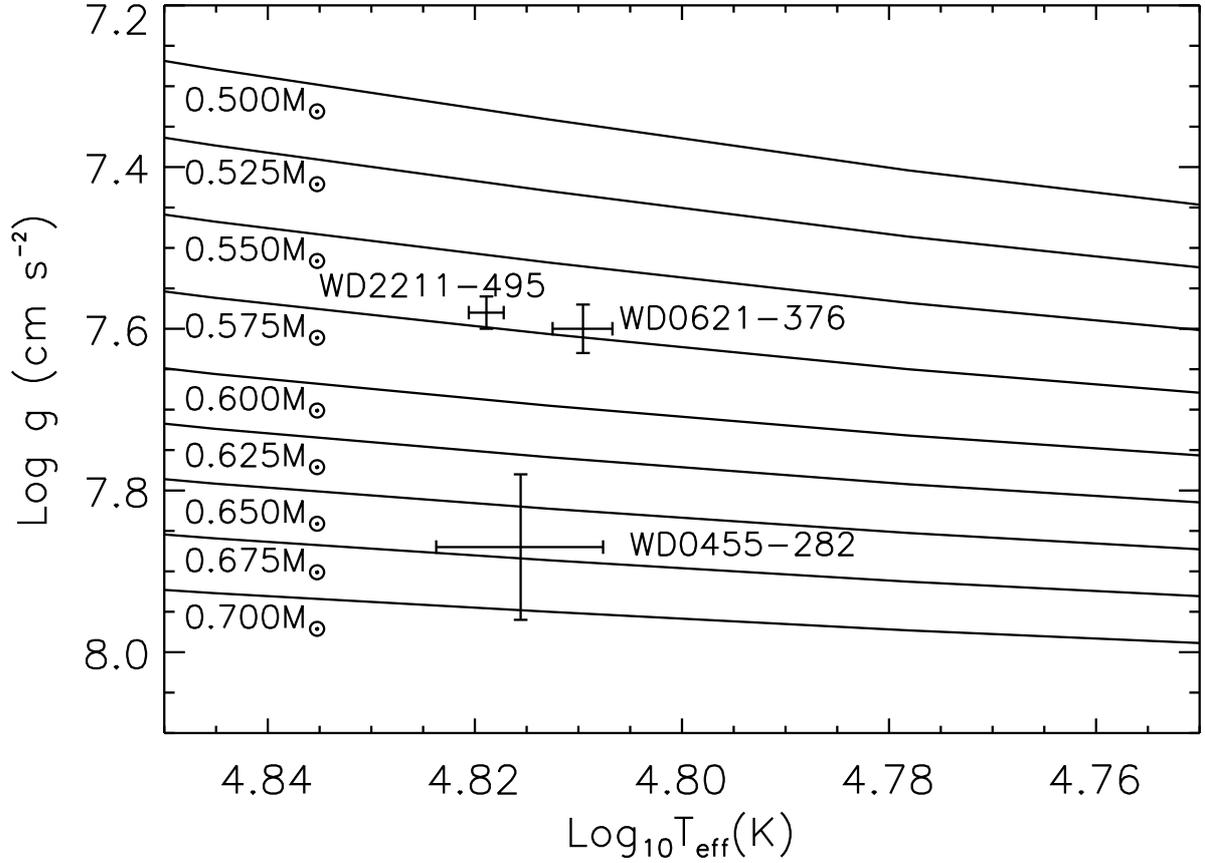}
\caption{Comparison of the measured values of 
$T_{\mathrm{eff}}$/log $g$ for the three white dwarfs with the montreal 
cooling curves for white dwarfs with masses ranging from 
0.5-0.7M$_\odot$. For comparison, the masses measured for WD0455-282, 
WD0621-376, and WD2211-495 were $0.589 \pm 0.145$, $0.731 \pm 0.053$,
and $0.575 \pm 0.028M_\odot$ respectively.}
\label{fig:mrcomp}
\end{centering}
\end{figure*}

As mentioned previously the high-precision astrometric data
from Gaia provides an excellent means with which to test the mass-radius 
relationship for white dwarf stars. In Figure \ref{fig:mrcomp} we have 
plotted our measured $T_{\mathrm{eff}}$/log $g$ values for the three white dwarfs, and 
the predicted $T_{\mathrm{eff}}$/log $g$ values from the Montreal cooling tables for 
white dwarfs with masses ranging from 0.5-0.7$M_\odot$. The cooling 
tables assume a thick H-envelope of $M_\mathrm{H}/M_{*}=1\times{10}^{-4}$,
where $M_{\mathrm{H}}$ is the mass of the H-layer, and $M_{*}$ is the mass
of the star. Using the best fitting synthetic spectrum and data from 
Gaia we measured the masses of WD0455-282, WD0621-376, and WD2211-495 to 
be $0.589 \pm 0.145$, $0.731 \pm 0.053$, and $0.575 \pm 0.028M_{\odot}$ respectively.
Comparatively, if we use the cooling tables to obtain the masses, we 
measure the masses to be $0.676 \pm 0.033$, $0.576 \pm 0.011$, and 
$0.572 \pm 0.007M_{\odot}$ for WD0455-282, WD0621-376, and WD2211-495 
respectively. For WD0455-282 and WD2211-495 the masses measured using 
Gaia data are in agreement within error with those measured using the 
cooling tables. However, there is a significant discrepancy in the case 
of WD0621-376. There are multiple potential explanations for this mass
discrepancy, which we now consider in turn. 

The first possibility concerns the accuracy of the parallax
used in the measurement. \cite{stassun2018a} found evidence of a $-82 \pm 33\mu$as 
systematic offset in the Gaia DR2 parallaxes. As the calculated mass is
dependent upon the radius, and, therefore, the parallax measurement, 
this shift will make the mass more uncertain. To check this, we applied
this offset to the Gaia DR2 parallax for WD0621-376 and recalculated the 
mass. This results in a mass of $0.740 \pm 0.053M_\odot$, showing that
the parallax measurement is not the cause of the discrepancy. Further
reinforcing this is the fact that WD0621-376 is not variable to within 
0.0029mag \citep{marinoni2016a}, ruling out uncertainties on the 
parallax due to variability. 
A second possibility concerns interstellar reddening, which can affect 
the UV flux and introduce uncertainties in the measured 
$T_{\mathrm{eff}}$ and log $g$. However, using the 3D dust maps of 
\cite{lallement2014a}, we find the extinction along the line of sight to 
WD0621-376 is negligible, and is unlikely to affect $T_{\mathrm{eff}}$ 
and log $g$.
A third possibility is the thickness of the H envelope assumed in the 
model calculation. Using detailed pre-white dwarf evolutionary sequences
\cite{romero2019a} showed that the thickness of this envelope decreases
as the total mass of the star increases, leading to a decrease in the
radius of the star of up to $\sim{12}$\%. However, this effect is too
small to explain the discrepant mass. 
Lastly, a fourth possibility is that the STIS spectrum is poorly calibrated. As 
an additional check, the V magnitude can be used to determine the mass 
of WD0621-376. This has the advantage of using optical rather than 
UV fluxes. In this case, the ratio $F_{\lambda}/H_{\lambda}$ seen in
Equation \ref{eqn:rad} is calculated by determining the flux 
$F_{\lambda}$ in the V-band: 
\begin{equation}
F_{\lambda}=F_{0}\times{10}^{-0.4V},
\end{equation}
where $V$ is the V-magnitude, $\lambda=5423$\AA, 
$F_{0}=3.804\times{10}^{-9}$ (see table 15 in \citealt{holberg2006a}), 
and $H_{\lambda}$ is the Eddington flux at $\lambda=5423$\AA. The mass
is then calculated using Equation \ref{eqn:mass}. By using 
$V=12.063\pm 0.020$ from Table \ref{table:starparam} we calculate the mass
to be $0.804 \pm 0.018M_{\odot}$. This is slightly larger than the 
value measured using the STIS data. This shows that the STIS calibration is
not the cause of the discrepant mass value.

The origin of this mass discrepancy is currently unknown, and requires
further study. This mass discrepancy could originate from the so-called 
Lyman/Balmer line problem described in \cite{barstow2003b}, where the
measured $T_{\mathrm{eff}}$/log $g$ values for a DA white dwarf differ 
depending on whether the Lyman or Balmer lines are used. We plan to 
explore this further in future work.

\section{Discussion: Discrepancies in WD0621-376}
The large difference between our value of log $g$ and that 
measured by B14 for WD0621-376 is a mystery. As well as the disagreement 
between our log $g$ measurement and that of B14, there is also 
disagreement with the value measured by \cite{gianninas2010a}. Using the 
Balmer H-lines, the authors measured log $g=7.12$. The disagreement 
between the optical and UV log $g$ measurements is not surprising as 
this has been seen in other stars as a consequence of the Lyman/Balmer line problem. 
A detailed discussion of the problem is beyond the scope of this paper. Despite the 
large disagreement between the various log $g$ measurements, we are 
confident that our measured $T_{\mathrm{eff}}$ for WD0621-376 is 
correct. This is because the ionization equilibria for O {\sc iv} and 
O {\sc v} are in excellent agreement with the observational data in 
Figures \ref{fig:threewd_oiv} and \ref{fig:threewd_ov} respectively. 
This can also be seen in the case of Fe {\sc v} and Fe {\sc vi} in 
Figures \ref{fig:threewd_fev} and \ref{fig:threewd_fevi} respectively.

Our measured log $g$ should at least be similar to that measured by B14
as we both used the same FUSE dataset (P10415010). However, the FUSE 
spectrum used in this work was constructed independently of B14. To see 
if using different spectra resulted in the discrepancy we measured 
$T_{\mathrm{eff}}$ and log $g$ of WD0621-376 using the same reduced FUSE 
spectrum as B14. For this measurement we used the $T_{\mathrm{eff}}$/log $g$ 
grid from the final iteration in this work for WD0621-376. Using the B14 
spectrum we measured $T_{\mathrm{eff}}=64800_{-521}^{+504}$K, and 
log $g=7.63_{-0.03}^{+0.03}$. Compared with the values measured in this
work of $T_{\mathrm{eff}}=64500_{-424}^{+437}$K and 
log $g=7.60_{-0.03}^{+0.03}$, the two results are in good agreement 
implying the log $g$ discrepancy does not originate from the dataset 
used.

The large uncertainty in log $g$ for WD0621-376 will also 
introduce large uncertainties in the measured abundances. To gauge this 
effect we re-measured the metal abundances for WD0621-376 using a model 
atmosphere with log $g=7.22$. The method used is the same as described 
in Section \ref{abnsec}. We set $T_{\mathrm{eff}}=63900$K as measured 
from the Lyman lines. In Table \ref{table:loggcomp} we list the measured 
abundances for WD0621-376 when using a model atmosphere with 
log $g=7.60$, and when using a model atmosphere with log $g=7.22$. We 
also list the \% difference between the two abundance measurements, 
calculated as 
\begin{equation}
\%\, \mathrm{Difference}=100\frac{(X-X_0)}{X_0},
\end{equation}
where $X$ is the abundance measured using the log $g=7.22$ model, and 
$X_0$ is the abundance measured using the log $g=7.60$ model. It can be
seen that the C, N, O, Fe, and Ni abundances are highly sensitive to the
different log $g$ values, with the abundance difference between the two 
models being 50.1, -53.7, -30.7, -46.5, and -42.7\% respectively. The 
Si, P, and S abundances are much less sensitive, with the abundance 
differences being 5.86, -0.19, and -17.1\% respectively.

The insensitivity of the Si abundance to log $g$ is surprising. In 
Section \ref{siabunsec} we noted that our Si abundance measurement for 
WD0621-376 was an order of magnitude larger than that measured by B14. 
Given the large difference between the log $g$ values measured in this
work and in B14, we would have expected at least some dependence upon
log $g$. Based on this fact, and the other factors hitherto discussed, 
it appears that the discrepancies described in this paper arises as a 
result of the stellar atmosphere grid used to make the measurements. 
Several differences exist between the model grids used in B14 and those 
used in the present work. Firstly, the models calculated in this work 
included more model atoms/ions than the B14 work, and covered a wider 
range of ionization states. Secondly, the Ni ions in this work used the 
photoionization cross sections calculated by \cite{preval2017a} using 
{\sc autostructure} \citep{badnell2011a}, whereas the B14 models used 
hydrogenic photoionization cross sections. Thirdly, the model grids used 
in the B14 study were not calculated based on an iterative scheme. In 
this case, the model grids calculated by B14 varied in 
$T_{\mathrm{eff}}$, log $g$, and metal abundance scaled by a 
multiplicative factor. Future work will be focused on a detailed 
exploration of all of these differences.

\begin{table}
\renewcommand{\arraystretch}{1.2}
\centering
\caption{Comparison of metal abundances measured for WD0621-376 using 
a model atmosphere with log $g=7.60$, and log $g=7.22$. Both models have
$T_{\mathrm{eff}}=64500$K. We have tabulated the \% difference between 
the measured abundances, defined in text.}
\begin{tabular}[H]{@{}lccc}
\hline
N(X)/N(H) & log $g=7.60$ & log $g=7.22$ & \% Difference \\
\hline
 N(C)/N(H) & $1.49_{-0.08}^{+0.08}\times{10}^{-6}$ & $2.24_{-0.09}^{+0.09}\times{10}^{-6}$ &  50.1 \\
 N(N)/N(H) & $6.45_{-0.12}^{+0.12}\times{10}^{-7}$ & $2.99_{-0.03}^{+0.04}\times{10}^{-7}$ & -53.7 \\
 N(O)/N(H) & $2.46_{-0.03}^{+0.03}\times{10}^{-6}$ & $1.70_{-0.03}^{+0.03}\times{10}^{-6}$ & -30.7 \\
N(Si)/N(H) & $8.75_{-0.19}^{+0.18}\times{10}^{-6}$ & $9.27_{-0.17}^{+0.17}\times{10}^{-6}$ &  5.86 \\
 N(P)/N(H) & $1.70_{-0.16}^{+0.16}\times{10}^{-7}$ & $1.69_{-0.16}^{+0.16}\times{10}^{-7}$ & -0.19 \\
 N(S)/N(H) & $1.05_{-0.06}^{+0.15}\times{10}^{-6}$ & $8.69_{-0.40}^{+0.39}\times{10}^{-7}$ & -17.1 \\
N(Fe)/N(H) & $1.84_{-0.04}^{+0.04}\times{10}^{-5}$ & $9.85_{-0.10}^{+0.10}\times{10}^{-6}$ & -46.5 \\
N(Ni)/N(H) & $4.78_{-0.08}^{+0.08}\times{10}^{-6}$ & $2.74_{-0.02}^{+0.02}\times{10}^{-6}$ & -42.7 \\
\hline
\end{tabular}
\label{table:loggcomp}
\renewcommand{\arraystretch}{1.0}
\end{table}

\section{Discussion: Interstellar \& Circumstellar absorption}\label{circumsec}
The work presented in this paper has been valuable in terms of our 
understanding of circumstellar absorption. In Section \ref{ismsec} we 
discussed our detections of circumstellar absorption features in 
WD0455-282, WD0621-376 and WD2211-495. As seen in previous works on 
circumstellar features, the absorption features detected in this work 
are all blue-shifted with respect to the photospheric velocity. We made 
several interesting observations in our study of the circumstellar 
absorption features, which we now discuss in turn.

\subsection{Si {\sc iii} detections}
Interestingly, while WD0455-282 and WD2211-495 showed signatures of 
circumstellar absorption, we made no detections of circumstellar 
absorption in the case of WD0621-376. However, the most interesting 
feature of this non-detection concerns the case of the Si {\sc iii} 
resonant feature with laboratory wavelength 1206.4995\AA. In the case of 
WD0621-376, we do not detect this feature at all, whereas for WD0455-282 
and WD2211-495, we do detect this feature. Because of the high 
temperatures and gravities of the stars studied, absorption from the 
Si {\sc iii} resonant transition in the photosphere will be very weak. 
This means that the feature shouldn't be detected in any of the stars 
studied. Indeed, in a study by \cite{preval2013a}, the authors also 
detected the Si {\sc iii} resonant transition in high-resolution STIS 
spectra of the hot white dwarf G191-B2B (see figure. 9 in the paper). The 
authors also detected a circumstellar component in the C {\sc iv} 
resonant doublet at 1548 and 1550\AA, and in the Si {\sc iv} resonant 
doublet at 1393 and 1402\AA\, (see figs. 5, 6, 10, and 11 in 
\citealt{preval2013a}). In the aforementioned works, Si {\sc iii} has 
not previously been considered as a circumstellar line. However, in the 
context of the stellar atmospheric ionization fractions, this line can 
only be circumstellar. Therefore, this implies that the resonant 
Si {\sc iii} line in hot white dwarfs could be a useful tracer of 
circumstellar absorption.

\subsection{Relation to the interstellar medium}
Previous studies such as those conducted by
\cite{bannister2003a}, \cite{dickinson2012b}, and \cite{preval2013a} 
have attempted to understand the origin of circumstellar absorption 
using velocity discrimination. For example, as mentioned previously in 
the case of G191-B2B, the C {\sc iv} doublet has both a photospheric and 
circumstellar component. \cite{preval2013a} measured the velocity the 
circumstellar components to be $8.26 \pm 0.18$ km s$^{-1}$ for the 
1548\AA\, line, and $8.30 \pm 0.15$ km s$^{-1}$ for the 1550\AA\, line. 
The authors also measured the average velocity of two ISM components 
along the line of sight to G191-B2B using low ionization-state resonant 
transitions commonly found in the ISM. They measured the velocities of 
these two components to be $19.4 \pm 0.03$ and
$8.64 \pm 0.03$ km s$^{-1}$ respectively. \cite{redfield2008a} showed 
that these two velocity components arise as a result of absorption from 
the LIC, and the Hyades cloud respectively. Interestingly, the velocity 
of the C {\sc iv} circumstellar components are very similar to the 
velocity measured for the Hyades cloud, tentatively implying a 
connection between the two.

In the present work, four ISM velocity components were 
detected along the line of sight to WD0455-282 with velocities of 
$v_{1}=-40.3 \pm 1.58$, $v_{2}=-32.1 \pm 3.04$, $v_{3}=16.7 \pm 1.85$, 
and $v_{4}=25.1 \pm 1.06$ km s$^{-1}$ respectively. For the Si {\sc iii} 
1206\AA\, line, we detected three circumstellar components with velocities of 
$-32.8 \pm 0.35$, $12.8 \pm 1.25$, and $23.4 \pm 0.25$ km s$^{-1}$ 
respectively. For the Si {\sc iv} 1393\AA\, line, we detected two 
circumstellar components with velocities of $17.3 \pm 4.89$ and 
$25.9 \pm 0.23$ km s$^{-1}$ respectively. Finally, for the Si {\sc iv} 
1402\AA\, line, we detected two circumstellar components with velocities
of $16.1 \pm 1.43$ and $25.3 \pm 0.36$ km s$^{-1}$ respectively.
The similarity of the circumstellar velocities to the $v_{2}$, $v_{3}$, 
and $v_{4}$ components is difficult to miss, and, like G191-B2B, implies 
some connection. 

In the case of WD2211-495 we detected two ISM components with velocities 
of $v_{1}=-9.91 \pm 1.39$ and $v_{2}=-2.18 \pm 2.43$ km s$^{-1}$ 
respectively. The velocity of the circumstellar component detected in 
the Si {\sc iii} 1206\AA\, line was measured to be 
$3.43 \pm 0.10$ km s$^{-1}$, and the velocity for the circumstellar 
component detected in the Si {\sc iv} 1393\AA\, line was measured to be 
$-1.11 \pm 0.56$ km s$^{-1}$. Both of these circumstellar velocities are 
similar to the ISM component $v_{2}$.

\subsection{Str\"{o}mgren sphere ionization}
\begin{figure}
\begin{centering}
\includegraphics[width=85mm]{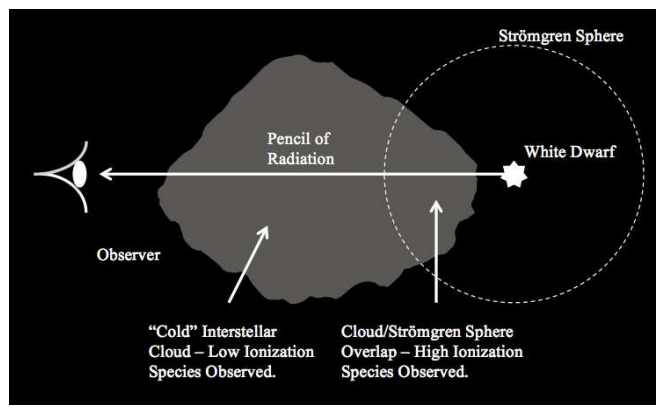}
\caption{Schematic of Str\"{o}mgren Sphere interacting with a cloud of 
gas. Radiation passing through the cloud outside the sphere produces low 
ionization state absorption features, while radiation passing through 
the cloud inside the sphere produces high ionization state absorption 
features.}
\label{fig:strom}
\end{centering}
\end{figure}

\cite{dickinson2012b} conducted a detailed survey of 23 white dwarf
stars with the aim of determining the origin of circumstellar absorption
features. One potential explanation touted in this work is that of the
Str\"{o}mgren Sphere. In this case, the UV radiation field of the white 
dwarf ionizes material within a certain radius of the object known as 
the Str\"{o}mgren radius \citep{stromgren1939a}. If a cloud in the ISM 
falls within this radius, it will produce lines from high ionization
state ions, while any material outside this radius will produce lines 
from neutral and low ionization state elements. To aid visualisation of 
this phenomenon, we have provided an illustration in Figure 
\ref{fig:strom}.

In the present work, we have identified circumstellar absorption in 
WD0455-282, and WD2211-495. We have also seen that the velocities of the
highly ionized circumstellar components are strikingly similar to those 
observed in the ISM lines with low ionization states. Furthermore, prior 
to the present work, circumstellar absorption has only been observed as 
a single velocity component. Now, for the first time, we have observed
circumstellar absorption with multiple velocity components along the 
line of sight to WD0455-282. These circumstellar velocities are all very 
similar to the average ISM velocities along the line of sight to 
WD0455-282. This is, perhaps, the strongest evidence to date that 
circumstellar absorption arises as a result of Str\"{o}mgren sphere 
ionization.

While Str\"{o}mgren sphere ionization explains how the low and high 
ionization state lines form, it does not explain where this material 
comes from. This material could potentially be inherent to the ISM, or
could even be inherent to the white dwarf itself, arising from the post
RGB pulsations of the object's progenitor state.

In \cite{dickinson2012b} the authors provide a table of predicted column 
densities for absorption features arising from Str\"{o}mgren sphere 
ionization. These column densities differ depending upon the transition
considered. The largest predicted column density occurs for the resonant
C {\sc iv} doublet at 1548 and 1550\AA. Unfortunately, the STIS data
used in this work stops short of the doublet. Therefore, for future
work, it would be useful to obtain high-resolution 
spectroscopy for WD0455-282 to observe the C {\sc iv} doublet. If the 
double component circumstellar feature is real, then this will also be 
observed in the C {\sc iv} doublet.

\section{Conclusions}
We have presented a detailed UV survey of three hot, DA white dwarf 
stars. Using data obtained with FUSE and HST, we have measured 
$T_{\mathrm{eff}}$, log $g$, and atmospheric compositions of these 
objects. In conjunction with measurements from the astrometric satellite 
Gaia, we have also directly measured the masses, radii, and 
gravitational redshift velocities of these stars. Upon 
comparison with the work of B14 we found our measurement 
of log $g$ (7.60) for WD0621-376 was much larger than the authors' value
(7.22). We tested the sensitivity of our abundance measurements for 
WD0621-376 to the different log $g$ values, and found the abundance
measurements differed greatly for C, N, O, Fe, and Ni ($>30$\%). More 
work needs to be done to understand the reason as to why our log $g$
measurement for WD0621-376 is much larger than that measured by B14.

We considered the question of why the three white dwarfs studied have 
metals in their atmospheres. To address this question, we compared the 
observed atmospheric abundances in the three white dwarfs to those 
predicted by radiative levitation theory as calculated by Ch94/95. We 
found that the observed Si abundance in all three white dwarfs was 
larger than predicted by an order of magnitude. We also found that the 
Fe abundance in WD0621-376 and WD2211-495 was larger than 
predicted by an order of magnitude. Given that the Ch94/95 abundances 
constitute a maximum possible supported abundance, this tentatively 
implies an external origin for metal pollution. However, it is not clear 
what this origin might be. This is further complicated by the interplay 
between radiative levitation, and gravitational diffusion. Therefore, to 
make any definitive progress on this topic, more sophisticated stellar 
atmospheric modelling including these processes is required.

In addition to our work on the photospheric features in the atmospheres 
of the three white dwarfs, we also conducted a survey of lines arising 
from absorption in the ISM. We made an interesting discovery along the 
line of sight to WD0455-282, where, for the first time, we detected 
multiple component circumstellar absorption in the Si {\sc iii} and 
Si {\sc iv} resonant lines. We also detected circumstellar absorption 
along the line of sight to WD2211-495 in the Si {\sc iv} 1393\AA\, line. 
We did not detect a circumstellar feature in the 1402\AA\, line. 
Interestingly, we made no detections of any circumstellar absorption 
along the line of sight to WD0621-376. We also could not detect the 
resonant Si {\sc iii} 1206\AA\, line along the line of sight to this 
white dwarf. This is in contrast to WD0455-282 and WD2211-495, where we 
detected very strong Si {\sc iii} absorption in both cases. This runs 
contrary to stellar atmosphere model predictions, where Si {\sc iii} is 
weakly populated in comparison to Si {\sc iv}. This implies that the 
Si {\sc iii} line can be used as a reliable tracer of circumstellar 
absorption along the line of sight to hot white dwarf stars.

\section*{Acknowledgements}
Based on observations from {\em Hubble Space Telescope}\/ collected at 
STScI, operated by the Associated Universities for Research in 
Astronomy, under contract to NASA. S.P.P., M.A.B., and M.B. greatfully 
acknowledge the financial support of the Leverhulme Foundation. N.R. 
greatfully acknowledges the support of a Royal Commission 1851 Research
Fellowshipt. J.B.H. and T.A. acknowledge support provided by NASA through 
grants from the Space Telescope Science Institute, which is operated by 
the Association of Universities for Research in Astronomy, Inc., under 
NASA contract NAS5-26555. JDB acknowledges support by the Science and 
Technology Facilities Council (STFC), UK. This research used the ALICE
High Performance Computing Facility at the University of Leicester.
We thank the anonymous referee for taking their time to review this 
paper. This work has made use of data from the European Space Agency (ESA) mission
{\it Gaia} (\url{https://www.cosmos.esa.int/gaia}), processed by the {\it Gaia}
Data Processing and Analysis Consortium (DPAC,
\url{https://www.cosmos.esa.int/web/gaia/dpac/consortium}). Funding for the DPAC
has been provided by national institutions, in particular the institutions
participating in the {\it Gaia} Multilateral Agreement.

\bibliographystyle{mnras}
\bibliography{simonpreval} 

\begin{thebibliography}{}
\makeatletter
\relax
\def\mn@urlcharsother{\let\do\@makeother \do\$\do\&\do\#\do\^\do\_\do\%\do\~}
\def\mn@doi{\begingroup\mn@urlcharsother \@ifnextchar [ {\mn@doi@}
  {\mn@doi@[]}}
\def\mn@doi@[#1]#2{\def\@tempa{#1}\ifx\@tempa\@empty \href
  {http://dx.doi.org/#2} {doi:#2}\else \href {http://dx.doi.org/#2} {#1}\fi
  \endgroup}
\def\mn@eprint#1#2{\mn@eprint@#1:#2::\@nil}
\def\mn@eprint@arXiv#1{\href {http://arxiv.org/abs/#1} {{\tt arXiv:#1}}}
\def\mn@eprint@dblp#1{\href {http://dblp.uni-trier.de/rec/bibtex/#1.xml}
  {dblp:#1}}
\def\mn@eprint@#1:#2:#3:#4\@nil{\def\@tempa {#1}\def\@tempb {#2}\def\@tempc
  {#3}\ifx \@tempc \@empty \let \@tempc \@tempb \let \@tempb \@tempa \fi \ifx
  \@tempb \@empty \def\@tempb {arXiv}\fi \@ifundefined
  {mn@eprint@\@tempb}{\@tempb:\@tempc}{\expandafter \expandafter \csname
  mn@eprint@\@tempb\endcsname \expandafter{\@tempc}}}

\bibitem[\protect\citeauthoryear{Arnaud}{Arnaud}{1996}]{arnaud1996a}
Arnaud K.~A.,  1996, in Jacoby G.~H.,  Barnes J.,  eds,  Astronomical Society
  of the Pacific Conference Series Vol. 101, Astronomical Data Analysis
  Software and Systems V. p.~17

\bibitem[\protect\citeauthoryear{Badnell}{Badnell}{2011}]{badnell2011a}
Badnell N.~R.,  2011, \mn@doi [Computer Physics Communications]
  {10.1016/j.cpc.2011.03.023}, 182, 1528

\bibitem[\protect\citeauthoryear{Bannister, Barstow, Holberg  \&
  Brahweiler}{Bannister et~al.}{2003}]{bannister2003a}
Bannister N.~P.,  Barstow M.~A.,  Holberg J.~B.,   Brahweiler F.~C.,  2003,
  \mn@doi [MNRAS] {10.1046/j.1365-8711.2003.06409.x}, 341, 477

\bibitem[\protect\citeauthoryear{Barstow, Good, Holberg, Hubeny, Bannister,
  Bruhweiler, Burleigh  \& Napiwotzki}{Barstow et~al.}{2003}]{barstow2003b}
Barstow M.~A.,  Good S.~A.,  Holberg J.~B.,  Hubeny I.,  Bannister N.~P.,
  Bruhweiler F.~C.,  Burleigh M.~R.,   Napiwotzki R.,  2003, \mn@doi [MNRAS]
  {10.1046/j.1365-8711.2003.06462.x}, 341, 870

\bibitem[\protect\citeauthoryear{Barstow, Barstow, Casewell, Holberg  \&
  Hubeny}{Barstow et~al.}{2014}]{barstow2014a}
Barstow M.~A.,  Barstow J.~K.,  Casewell S.~L.,  Holberg J.~B.,   Hubeny I.,
  2014, \mn@doi [MNRAS] {10.1093/mnras/stu216}, 440, 1607

\bibitem[\protect\citeauthoryear{Bergeron et~al.,}{Bergeron
  et~al.}{2011}]{bergeron2011a}
Bergeron P.,  et~al., 2011, \mn@doi [ApJ] {10.1088/0004-637X/737/1/28}, 737, 28

\bibitem[\protect\citeauthoryear{Chayer, Fontaine  \& Wesemael}{Chayer
  et~al.}{2002a}]{chayer1995a}
Chayer P.,  Fontaine G.,   Wesemael F.,  2002a, \mn@doi [ApJS]
  {10.1086/192184}, 99, 189

\bibitem[\protect\citeauthoryear{Chayer, LeBlanc, Fontaine, Wesemael, Michaud
  \& Vennes}{Chayer et~al.}{2002b}]{chayer1994a}
Chayer P.,  LeBlanc F.,  Fontaine G.,  Wesemael F.,  Michaud G.,   Vennes S.,
  2002b, \mn@doi [ApJ] {10.1086/187657}, 436, L161

\bibitem[\protect\citeauthoryear{Chayer, Vennes, Pradhan, Thejll, Beauchamp,
  Fontaine  \& Wesemael}{Chayer et~al.}{2002c}]{chayer1995b}
Chayer P.,  Vennes S.,  Pradhan A.~K.,  Thejll P.,  Beauchamp A.,  Fontaine G.,
    Wesemael F.,  2002c, \mn@doi [ApJ] {10.1086/176494}, 454, 429

\bibitem[\protect\citeauthoryear{Croll et~al.,}{Croll
  et~al.}{2015}]{croll2017a}
Croll B.,  et~al., 2015, \mn@doi [ApJ] {10.3847/1538-4357/836/1/82}, 836, 82

\bibitem[\protect\citeauthoryear{Dickinson, Barstow  \& Hubeny}{Dickinson
  et~al.}{2012}]{dickinson2012b}
Dickinson N.~J.,  Barstow M.~A.,   Hubeny I.,  2012, \mn@doi [MNRAS]
  {10.1111/j.1365-2966.2012.20545.x}, 421, 3222

\bibitem[\protect\citeauthoryear{G{\"{a}}nsicke, Marsh, Southworth  \&
  Rebassa-Mansergas}{G{\"{a}}nsicke et~al.}{2006}]{gansicke2006a}
G{\"{a}}nsicke B.~T.,  Marsh T.~R.,  Southworth J.,   Rebassa-Mansergas A.,
  2006, \mn@doi [Science] {10.1126/science.1135033}, 314, 1908

\bibitem[\protect\citeauthoryear{G{\"{a}}nsicke et~al.,}{G{\"{a}}nsicke
  et~al.}{2016}]{gansicke2016a}
G{\"{a}}nsicke B.~T.,  et~al., 2016, \mn@doi [ApJ]
  {10.3847/2041-8205/818/1/l7}, 818, L7

\bibitem[\protect\citeauthoryear{Gianninas, Bergeron, Dupuis  \&
  Ruiz}{Gianninas et~al.}{2010}]{gianninas2010a}
Gianninas A.,  Bergeron P.,  Dupuis J.,   Ruiz M.~T.,  2010, \mn@doi [ApJ]
  {10.1088/0004-637X/720/1/581}, 720, 581

\bibitem[\protect\citeauthoryear{Henden, Templeton, Terrell, Smith, Levine  \&
  Welch}{Henden et~al.}{2016}]{henden2016a}
Henden A.,  Templeton M.,  Terrell D.,  Smith T.,  Levine S.,   Welch D.,
  2016, VizieR Online Data Catalog, p. II/336

\bibitem[\protect\citeauthoryear{Holberg \& Bergeron}{Holberg \&
  Bergeron}{2006}]{holberg2006a}
Holberg J.~B.,  Bergeron P.,  2006, \mn@doi [AJ] {10.1086/505938}, 132, 1221

\bibitem[\protect\citeauthoryear{Holberg, Barstow, Bruhweiler, Cruise  \&
  Penny}{Holberg et~al.}{2002}]{holberg1998a}
Holberg J.~B.,  Barstow M.~A.,  Bruhweiler F.~C.,  Cruise A.~M.,   Penny A.~J.,
   2002, \mn@doi [ApJ] {10.1086/305489}, 497, 935

\bibitem[\protect\citeauthoryear{Hu et~al.,}{Hu et~al.}{2018}]{hu2019a}
Hu J.,  et~al., 2018, arXiv e-prints

\bibitem[\protect\citeauthoryear{Hubeny}{Hubeny}{1988}]{hubeny1988a}
Hubeny I.,  1988, \mn@doi [Computer Physics Communications]
  {10.1016/0010-4655(88)90177-4}, 52, 103

\bibitem[\protect\citeauthoryear{Hubeny \& Lanz}{Hubeny \&
  Lanz}{2002}]{hubeny1995a}
Hubeny I.,  Lanz T.,  2002, \mn@doi [ApJ] {10.1086/175226}, 439, 875

\bibitem[\protect\citeauthoryear{Hubeny \& Lanz}{Hubeny \&
  Lanz}{2011}]{hubeny2011a}
Hubeny I.,  Lanz T.,  2011, in Astrophysics Source Code Library, record
  ascl:1109.022. p.~9022

\bibitem[\protect\citeauthoryear{Iben \& Tutukov}{Iben \&
  Tutukov}{1997}]{iben1997a}
Iben I.,  Tutukov A.,  1997, $\backslash$Skytel, 94, 36

\bibitem[\protect\citeauthoryear{Kawka \& Vennes}{Kawka \&
  Vennes}{2016}]{kawka2016a}
Kawka A.,  Vennes S.,  2016, \mn@doi [MNRAS] {10.1093/mnras/stw383}, 458, 325

\bibitem[\protect\citeauthoryear{Koester, G{\"{a}}nsicke  \& Farihi}{Koester
  et~al.}{2014}]{koester2014a}
Koester D.,  G{\"{a}}nsicke B.~T.,   Farihi J.,  2014, \mn@doi [Astronomy {\&}
  Astrophysics] {10.1051/0004-6361/201423691}, 566, A34

\bibitem[\protect\citeauthoryear{Kowalski \& Saumon}{Kowalski \&
  Saumon}{2006}]{kowalski2006a}
Kowalski P.~M.,  Saumon D.,  2006, \mn@doi [ApJ] {10.1086/509723}, 651, L137

\bibitem[\protect\citeauthoryear{Kurucz}{Kurucz}{1992}]{kurucz1992a}
Kurucz K.,  1992, \mn@doi [Revista Mexicana de Astronomia y Astrofisica]
  {1992RMxAA..23...45K}, 23, 45

\bibitem[\protect\citeauthoryear{Lallement, Vergely, Valette, Puspitarini, Eyer
   \& Casagrande}{Lallement et~al.}{2014}]{lallement2014a}
Lallement R.,  Vergely J.-L.,  Valette B.,  Puspitarini L.,  Eyer L.,
  Casagrande L.,  2014, \mn@doi [A{\&}A] {10.1051/0004-6361/201322032}, 561,
  A91

\bibitem[\protect\citeauthoryear{Landenberger-Schuh}{Landenberger-Schuh}{2005}]{schuh2005a}
Landenberger-Schuh S.,  2005, PhD thesis, Institut f{\"{u}}r Astronomie und
  Astrophysik Tuebingen, Sand 1, D-72076 Tuebingen, Germany;
  Universit{\"{a}}tssternwarte G{\"{o}}ttingen, Geismarlandstrasse 11, D-37083
  G{\"{o}}ttingen, Germany
  {\textless}EMAIL{\textgreater}schuh@astro.physik.uni-goettingen.de{\textless}/EMAIL{\textgreater},
  \url {http://adsabs.harvard.edu/abs/2005PhDT.........1L}

\bibitem[\protect\citeauthoryear{Lemke}{Lemke}{2003}]{lemke1997a}
Lemke M.,  2003, \mn@doi [Astronomy and Astrophysics Supplement Series]
  {10.1051/aas:1997134}, 122, 285

\bibitem[\protect\citeauthoryear{Manser}{Manser}{2019}]{manser2019a}
Manser C.~J.,  2019, \mn@doi [Science] {10.1126/science.aat5330}, 364, 66

\bibitem[\protect\citeauthoryear{Marinoni et~al.,}{Marinoni
  et~al.}{2016}]{marinoni2016a}
Marinoni S.,  et~al., 2016, \mn@doi [MNRAS] {10.1093/mnras/stw1886}, 462, 3616

\bibitem[\protect\citeauthoryear{Markwardt}{Markwardt}{2009}]{markwardt2009a}
Markwardt C.~B.,  2009, in Bohlender D.~A.,  Durand D.,   Dowler P.,  eds,
  Astronomical Society of the Pacific Conference Series Vol. 411, Astronomical
  Data Analysis Software and Systems XVIII. p.~251 (\mn@eprint {arXiv}
  {0902.2850}), \url {http://arxiv.org/abs/0902.2850}

\bibitem[\protect\citeauthoryear{Martin \& Baltimore}{Martin \&
  Baltimore}{2012}]{hernandez2012a}
Martin S.,  Baltimore D.,  2012, {Space Telescope Imaging Spectrograph
  Instrument Handbook for Cycle 21}.
No. December

\bibitem[\protect\citeauthoryear{Moos et~al.,}{Moos et~al.}{2002}]{moos2000a}
Moos H.~W.,  et~al., 2002, \mn@doi [ApJ] {10.1086/312795}, 538, L1

\bibitem[\protect\citeauthoryear{Paquette, Pelletier, Fontaine  \&
  Michaud}{Paquette et~al.}{2002}]{paquette1986a}
Paquette C.,  Pelletier C.,  Fontaine G.,   Michaud G.,  2002, \mn@doi [ApJS]
  {10.1086/191112}, 61, 197

\bibitem[\protect\citeauthoryear{Preval, Barstow, Holberg  \& Dickinson}{Preval
  et~al.}{2013}]{preval2013a}
Preval S.~P.,  Barstow M.~A.,  Holberg J.~B.,   Dickinson N.~J.,  2013, \mn@doi
  [MNRAS] {10.1093/mnras/stt1604}, 436, 659

\bibitem[\protect\citeauthoryear{Preval, Barstow, Badnell, Hubeny  \&
  Holberg}{Preval et~al.}{2017}]{preval2017a}
Preval S.~P.,  Barstow M.~A.,  Badnell N.~R.,  Hubeny I.,   Holberg J.~B.,
  2017, \mn@doi [MNRAS] {10.1093/mnras/stw2800}, 465, 269

\bibitem[\protect\citeauthoryear{Rappaport, Gary, Kaye, Vanderburg, Croll,
  Benni  \& Foote}{Rappaport et~al.}{2016}]{rappaport2016a}
Rappaport S.,  Gary B.~L.,  Kaye T.,  Vanderburg A.,  Croll B.,  Benni P.,
  Foote J.,  2016, \mn@doi [MNRAS] {10.1093/mnras/stw612}, 458, 3904

\bibitem[\protect\citeauthoryear{Redfield \& Linsky}{Redfield \&
  Linsky}{2008}]{redfield2008a}
Redfield S.,  Linsky J.~L.,  2008, \mn@doi [ApJ] {10.1086/524002}, 673, 283

\bibitem[\protect\citeauthoryear{Romero, Kepler, Joyce, Lauffer  \&
  C{\'{o}}rsico}{Romero et~al.}{2019}]{romero2019a}
Romero A.~D.,  Kepler S.~O.,  Joyce S. R.~G.,  Lauffer G.~R.,   C{\'{o}}rsico
  A.~H.,  2019, \mn@doi [MNRAS] {10.1093/mnras/stz160}, 484, 2711

\bibitem[\protect\citeauthoryear{Seaton, Yan, Mihalas  \& Pradhan}{Seaton
  et~al.}{2014}]{seaton1994a}
Seaton M.~J.,  Yan Y.,  Mihalas D.,   Pradhan A.~K.,  2014, \mn@doi [MNRAS]
  {10.1093/mnras/266.4.805}, 266, 805

\bibitem[\protect\citeauthoryear{Stassun \& Torres}{Stassun \&
  Torres}{2018}]{stassun2018a}
Stassun K.~G.,  Torres G.,  2018, \mn@doi [ApJ] {10.3847/1538-4357/aacafc},
  862, 61

\bibitem[\protect\citeauthoryear{Str{\"{o}}mgren}{Str{\"{o}}mgren}{1939}]{stromgren1939a}
Str{\"{o}}mgren B.,  1939, \mn@doi [ApJ] {10.1086/144074}, 89, 526

\bibitem[\protect\citeauthoryear{Tremblay \& Bergeron}{Tremblay \&
  Bergeron}{2009}]{tremblay2009a}
Tremblay P.~E.,  Bergeron P.,  2009, \mn@doi [ApJ]
  {10.1088/0004-637X/696/2/1755}, 696, 1755

\bibitem[\protect\citeauthoryear{Tremblay, Bergeron  \& Gianninas}{Tremblay
  et~al.}{2011}]{tremblay2011a}
Tremblay P.~E.,  Bergeron P.,   Gianninas A.,  2011, \mn@doi [ApJ]
  {10.1088/0004-637X/730/2/128}, 730, 128

\bibitem[\protect\citeauthoryear{Vanderburg et~al.,}{Vanderburg
  et~al.}{2015}]{vanderburg2015a}
Vanderburg A.,  et~al., 2015, \mn@doi [Nature] {10.1038/nature15527}, 526, 546

\bibitem[\protect\citeauthoryear{Wilson, G{\"{a}}nsicke, Farihi  \&
  Koester}{Wilson et~al.}{2016}]{wilson2016a}
Wilson D.~J.,  G{\"{a}}nsicke B.~T.,  Farihi J.,   Koester D.,  2016, \mn@doi
  [MNRAS] {10.1093/mnras/stw844}, 459, 3282

\bibitem[\protect\citeauthoryear{Xu, Jura, Koester, Klein  \& Zuckerman}{Xu
  et~al.}{2014}]{xu2014a}
Xu S.,  Jura M.,  Koester D.,  Klein B.,   Zuckerman B.,  2014, \mn@doi [ApJ]
  {10.1088/0004-637X/783/2/79}, 783, 79

\bibitem[\protect\citeauthoryear{Xu, Jura, Dufour  \& Zuckerman}{Xu
  et~al.}{2016}]{xu2016a}
Xu S.,  Jura M.,  Dufour P.,   Zuckerman B.,  2016, \mn@doi [ApJ]
  {10.3847/2041-8205/816/2/l22}, 816, L22

\bibitem[\protect\citeauthoryear{Zuckerman, Koester, Dufour, Melis, Klein  \&
  Jura}{Zuckerman et~al.}{2011}]{zuckerman2011a}
Zuckerman B.,  Koester D.,  Dufour P.,  Melis C.,  Klein B.,   Jura M.,  2011,
  \mn@doi [ApJ] {10.1088/0004-637X/739/2/101}, 739, 101

\makeatother
\end{thebibliography}

\appendix
\section{Gaussian parameterisation}
To parameterise the non-photospheric absorption features we used a 
Gaussian profile. Because there can be a large number of absorbers along 
the line of sight, the profile can often become saturated. We account 
for this by using the full expression
\begin{equation} 
F_{\lambda}=\prod_{i=1}^{N}{F_{i}} = F_{0}\prod_{i=1}^{N}\exp{\left[-P_{0,i}\phi\right]},
\end{equation}
where $F_{0}$ is the continuum flux, $P_{0,i}$ is the absorption 
strength for the $i$th Gaussian profile, and $\phi$ is the normalised 
Gaussian function. The product is calculated out over $N$ absorption 
profiles. $\phi$ is then written as
\begin{equation}
\phi_{\lambda}=\frac{1}{P_{1,i}\sqrt\pi}\exp\left[-\frac{(\lambda-P_{2,i})^2}{P_{1,i}^2}\right],
\end{equation}
where $P_{1,i}$ is the Doppler width of the $i$th Gaussian, and 
$P_{2,i}$ is the centroid wavelength of the $i$th Gaussian. Details on 
the fitting procedure can be found in the text.

\bsp	
\label{lastpage}
\end{document}